\PassOptionsToPackage{hyperfootnotes=false}{hyperref}
\documentclass[11pt]{article}
\usepackage[final]{acl}

\usepackage{times}
\usepackage{latexsym}
\usepackage{enumitem}
\usepackage{bm}

\usepackage[T1]{fontenc}

\usepackage[utf8]{inputenc}

\usepackage{microtype}

\usepackage{inconsolata}
\usepackage{float}

\usepackage{graphicx}
\usepackage{amsmath}
\usepackage{amssymb}
\usepackage{booktabs}
\usepackage{longtable}
\usepackage{comment}
\usepackage{xcolor}

\usepackage{tikz}
\usetikzlibrary{arrows.meta, bending, positioning, matrix}

\newcommand{\shownotes}{1} 
\ifnum \shownotes = 1
	\newcommand{\authnote}[2]{{$\ll$\textsf{\footnotesize #1 notes: #2}$\gg$}}
\else
	\newcommand{\authnote}[2]{}
\fi

\title{Two-Sided State-Space Models for Sequential Recommendation with Non-Random Multimodal Review Feedback}

\author{
Ziwen Pan\thanks{Equal contribution. Code and reproducibility materials are available at \href{https://github.com/CausalMLResearch/TS-SSM}{here} under the MIT License.}
\quad
Zihan Liang\footnotemark[1]
\quad
Ruoxuan Xiong \\
Emory University, Atlanta, USA \\
\texttt{\{ziwen.pan, zihan.liang, ruoxuan.xiong\}@emory.edu}
}

\begin{document}
\maketitle
\begin{abstract}
Two-sided digital platforms are inherently dynamic: user preferences shift, item popularity evolves, and reviews both reflect and drive these changes. Yet most sequential recommendation systems treat reviews as passive signals for updating user states, leaving two aspects underexplored. First, review generation is nonrandom, depending on evolving latent states of both users and items. Second, reviews can reshape item states, induce spillover across related items, and influence future user decisions. To address these gaps, we propose a two-sided state-space model (TS-SSM) for event-conditioned sequential recommendation. TS-SSM consists of three components: (1) a modality-missing-not-at-random fusion module that encodes review content and informative observation patterns; (2) user-state evolution with temporal variation and local graph message passing that uses related item states to refine user preferences; and (3) item-state evolution with asymmetric carryover of positive and negative review feedback. In experiments across six Amazon categories, TS-SSM increases Recall@20 over BSARec by 14.8\%--18.8\% and exceeds HM4SR by 11.7\% on average. On Goodreads Fantasy, Recall@20 improves HM4SR from .5191 to .5847. 
Ablations highlight distinct contributions of observation patterns, local propagation, and item dynamics.
\end{abstract}

\section{Introduction}
\label{sec:intro}

Two-sided digital platforms are ubiquitous, spanning settings such as e-commerce marketplaces and service platforms. Both users and items evolve over time: user preferences shift with accumulated experience~\citep{koren2010collaborative}, while item perception can change as public feedback accumulates.

User reviews play a central role in these dynamics. They capture user experience through ratings, text, and images, providing rich signals for modeling user preferences. Review-aware methods use review text to learn user and item representations~\citep{mcauley2013hidden,zheng2017joint,wu2019reviews}, while recent text-aware sequential recommendation systems use item textual attributes or semantic item embeddings~\citep{li2023text,hu2024said}. However, these approaches largely treat the item side as static.

\begin{figure}[ht!]
    \centering
    \includegraphics[width=\columnwidth]{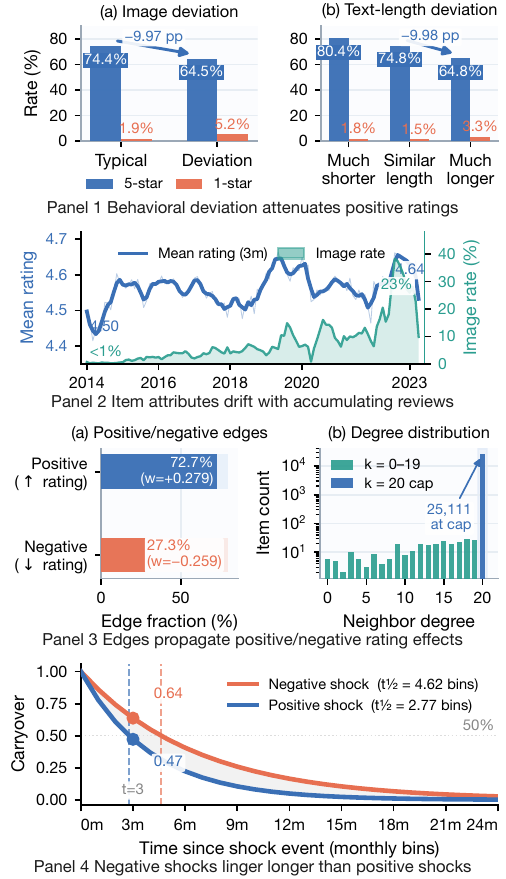}
    \caption{TS-SSM motivations: (\textbf{1})~predictive observation patterns, as behavioral deviations accompany fewer 5-star and more 1-star reviews; (\textbf{2})~item-attribute drift across 112 train-aligned bins (monthly through March 2023; April--August merged); (\textbf{3})~sign-aware review propagation over a dense unsigned co-occurrence graph; and (\textbf{4})~distinct half-lives from default negative/positive decay initializations ($t_{1/2}=4.62/2.77$ bins).}
    \label{fig:deviation_pattern}
\end{figure}

This leaves two important aspects underexplored. First, how are reviews generated? Review generation is neither random nor passive: whether a user leaves a review, and how it is expressed through optional modalities, depend on evolving latent user and item states. As shown in Figure~\ref{fig:deviation_pattern} using Amazon review data~\citep{hou2024bridging}, unusual image uploads or abrupt changes in text length are systematically associated with shifts in ratings, indicating that the observation pattern itself is informative.

Second, how do reviews propagate through item-side dynamics? Items evolve as feedback accumulates, so later users encounter items whose reputations are shaped by prior reviews. As summarized in Figure~\ref{fig:deviation_pattern}, item attributes drift over time, and evidence from related items can inform recommendations through a local graph.

We therefore propose a two-sided state-space model (TS-SSM) for sequential recommendation. TS-SSM maintains dynamic user and item states and models how reviews reflect and reshape them, with review text integrated throughout the model.

TS-SSM makes three methodological contributions. First, we develop a modality-missing-not-at-random review-fusion module, which uses (i) the presence or absence of each modality, such as text or images, and (ii) deviations from a user's historical expression patterns as signals of evolving user and item states. This builds on prior work on non-random modality missingness in multimodal records~\citep{liang-etal-2025-causal, liang-etal-2026-learning}, but extends it to jointly updating user and item states.

Second, we develop user-state evolution that captures temporal variation and uses local graph message passing to refine user preferences. The key idea is to propagate review-induced preference information through item relationships, allowing related items to inform the user-state update.

Third, we develop item-state evolution that captures how reviews reshape item states over time. In particular, prior reviews enter through a carryover mechanism with separate decay rates for positive and negative feedback, allowing their influence on the item state to persist asymmetrically over time.

We conduct extensive experiments across multiple domains and find that TS-SSM consistently outperforms existing methods. Across six Amazon categories~\citep{hou2024bridging}, it improves Recall@20 over BSARec by $14.8\%$--$18.8\%$ and beats the strongest recent baseline, HM4SR, by $11.7\%$ on average. On the Goodreads Book Graph's Fantasy subset~\citep{wan2018monotonic,wan2019spoiler}, it improves over HM4SR by $12.6\%$. Replacing dynamic event-level review text with static item-level text lowers Recall@20 by $2.67\%$ versus the full model; removing text lowers it by $6.30\%$. Appendices~\ref{app:goodreads}, \ref{app:ablation_extended}, and~\ref{app:encoder_sensitivity} report cross-platform results and sensitivity analyses.

\section{Problem Formulation}
\label{sec:problem}

We study sequential recommendation on a bipartite user--item system. Let $\mathcal{U}$ and $\mathcal{I}$ denote user and item sets, respectively. For an event time $t$, let $b(t)\in\{1,\ldots,T\}$ denote its coarse time bin; Appendix~\ref{app:impl:groups} specifies the train-aligned bin construction.

\paragraph{Bipartite states.}
At time $t$, the system state comprises the latent states of all users and items:
\begin{equation*}
\mathbf{S}_t = (\mathbf{Z}_t, \mathbf{V}_t),~~ \mathbf{Z}_t = [Z_{u,t}]_{u \in \mathcal{U}},~~ \mathbf{V}_t = [V_{i,t}]_{i \in \mathcal{I}}.
\end{equation*}

Here, $Z_{u,t}$ and $V_{i,t}$ denote the latent states of user $u$ and item $i$, respectively, and we later use $z_{u,t}$ and $v_{i,t}$ to denote their realizations.

\paragraph{Events.}
An event occurs when user $u$ leaves a review for item $i$. We represent the event as
\begin{equation*}
e_{u,i}=\left(\tau_{u,i},\mathcal O_{u,i}\right),
\end{equation*}
where $\tau_{u,i}$ denotes the event time and 
\begin{equation*}
\mathcal{O}_{u,i}=\left(M_{u,i}, A_{u,i}, r_{u,i}, \kappa_{u,i}\right),
\end{equation*}
denotes the observed review. Here, $M_{u,i}$ contains the multimodal review content, e.g., title, text, and images; $A_{u,i}$ is a binary vector indicating which modalities are observed; $r_{u,i}$ is the review rating; and $\kappa_{u,i}$ contains additional numeric review cues. 

Consider an event $e_{u,i}$ occurring at time $t=\tau_{u,i}$.  Review observation can be state-dependent. Both the modality-availability pattern $A_{u,i}$ and the observed review content may depend on the pre-event user and item states $Z_{u,t^-}$ and $V_{i,t^-}$, where $t^-$ denotes the time immediately before $t$. 

We define the corresponding pre-event global, user, and item histories as
\begin{align*}
\mathcal H_{t^-}&=\{e_{u',i'}:\tau_{u',i'} < t,\ \forall u',i'\}\\
\mathcal H_{u,t^-}&=\{e_{u,i'}:\tau_{u,i'} < t,\ \forall i'\}\\
\mathcal H_{i,t^-}&=\{e_{u',i}:\tau_{u',i}<t,\ \forall u'\}.
\end{align*}
The post-event history is $\mathcal H_t=\mathcal H_{t^-}\cup\{e_{u,i}\}$. 

\paragraph{Event representation and state transitions.}
At time $t = \tau_{u,i}$, we represent the event $e_{u,i}$ as
\begin{equation}\label{eqn:event-representation}
\Phi_{u,i,t}=f_{\mathrm{obs}}\left(\tau_{u,i},\mathcal O_{u,i},\phi_i,t\right),
\end{equation}
where $\phi_i$ is a learned representation of static item features, such as its description. 
The collection of available event representations at time $t$ is
\begin{equation}
\mathbf{\Phi}_t=\left\{\Phi_{u',i',\tau_{u',i'}}:\tau_{u',i'}\leq t,\ \forall u',i'\right\},
\end{equation}
which includes $\Phi_{u,i,t}$ of the current event $e_{u,i}$.

Upon the arrival of $e_{u,i}$, the user and item states evolve according to
\begin{align}
Z_{u,t}&=f^{U}\left(Z_{u,t^-}, \mathbf{\Phi}_t,  X_{u,t}\right) + \varepsilon^{U}_{u,t}\label{eqn:user-evolution}
\\
V_{i,t}&=f^{I}\left(V_{i,t^-}, \mathbf{\Phi}_t, Y_{i,t}\right) + \varepsilon^{I}_{i,t},\label{eqn:item-evolution}
\end{align}
where $f^U$ and $f^I$ are transition functions, and $X_{u,t}$ and $Y_{i,t}$ denote global temporal contexts for users (e.g., seasonal preferences) and items (e.g., platform-wide or macroeconomic shifts).

\paragraph{Ranking-based evaluation.}  At event time $t=\tau_{u,i}$, we use the pre-event history $\mathcal H_{t^-}$ together with the current review $\mathcal O_{u,i}$ to construct the post-event user state $z_{u,t}$ and the candidate item states $v_{j,t}$, for $j\in\mathcal I$. Then we compute a score
\begin{equation}
y_{u,j,t}=z_{u,t}^{\top}v_{j,t}+\mu_{j,t},\label{eqn:score}
\end{equation}
where $\mu_{j,t}$ is a learned dynamic item-specific bias, specified in Section~\ref{sec:method:training}. 

We rank the observed item $i$ against all items $j\in\mathcal I$ using $y_{u,j,t}$ and evaluate the rank of $i$ using Recall@$K$ and NDCG@$K$. Higher values indicate better ranking performance.\footnote{For this evaluation metric, we omit a user-specific bias from Equation~\eqref{eqn:score}, as it is constant across candidate items for a given user and does not affect the ranking.}

\section{Method}\label{sec:method}
\begin{figure*}[t]
    \centering
    \includegraphics[width=\textwidth]{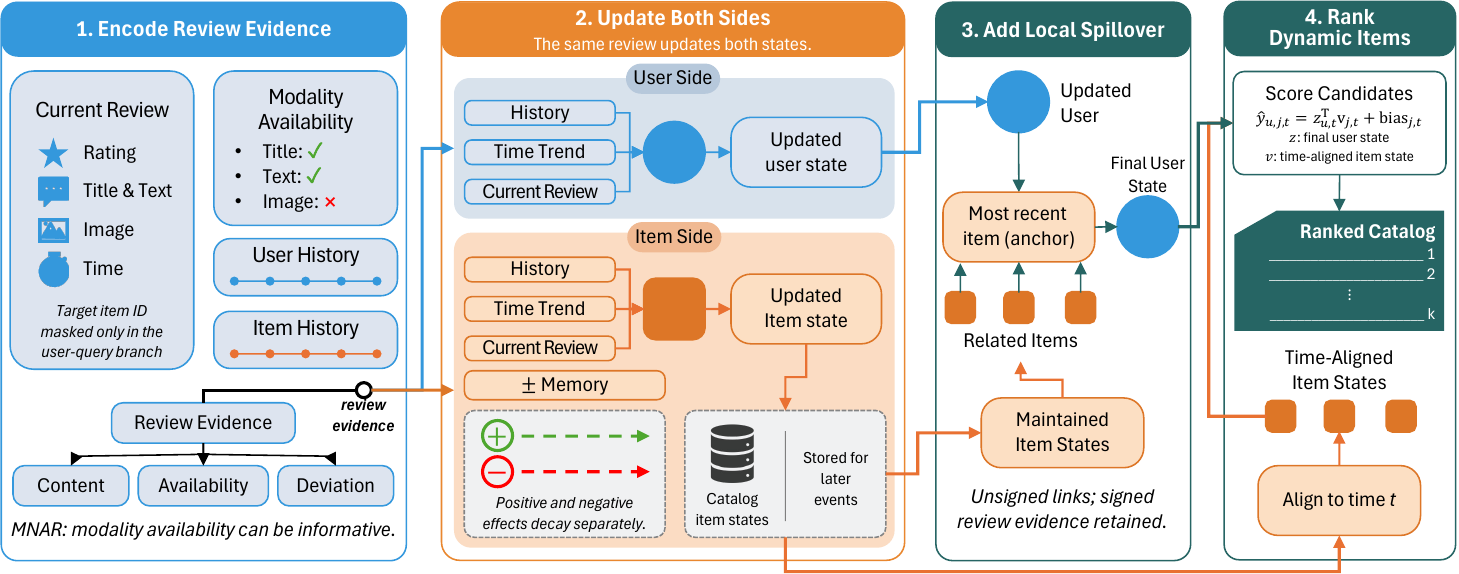}
    \caption{Overview of \textsc{TS-SSM}. {Stage 1} encodes multimodal content, availability patterns, and historical deviations; {Stage 2} updates user/item states with asymmetric item-side carryover; {Stage 3} refines user state by propagating review information over an item graph; and {Stage 4} aligns item states to query time and ranks candidates.}
    \label{fig:framework}
\end{figure*}

We propose TS-SSM, a two-sided state-space model for event-conditioned sequential recommendation. For each review event $e_{u,i}$, it encodes the review relative to user and item histories (Section~\ref{sec:method:deviation}), updates both states (Sections~\ref{sec:method:hsd} and~\ref{subsec:item-evolution}), and aligns candidate-item states to query time $t$ for ranking (Section~\ref{sec:method:training}). Figures~\ref{fig:framework} and~\ref{fig:causal_graph} summarize its architecture and state evolution.

\subsection{Multimodal Review Encoding}
\label{sec:method:deviation}

Our encoding procedure extracts two complementary sources of information from each review event: (i) the observed review record, including the presence or absence of each modality and its observed content, and (ii) deviations of the current review from historical user and item expression patterns. We describe these two components below.

\smallskip
\noindent\textit{Review-record encoding.}
We first specify the event observation encoder $f_{\mathrm{obs}}$ in Equation~\eqref{eqn:event-representation} to obtain the representation $\Phi_{u,i,t}$. The encoder incorporates three types of information. First, we encode the review content and explicitly account for modality availability through $f_{\mathrm{mod}}\left(\operatorname{Enc}(M_{u,i}),A_{u,i}\right)$.

Second, we encode the review rating and additional numeric cues through $f_{\mathrm{num}}\left(r_{u,i},\kappa_{u,i}\right)$.

Third, we incorporate learned representations $\phi_i$, $\phi_t$, and $\phi_{A_{u,i}}$ for static item features (e.g., item description), query-time features, and the modality-availability pattern, respectively. Notably, $\phi_{A_{u,i}}$ allows the observation pattern itself to provide information beyond its role in multimodal fusion.

Combining these components, we instantiate $f_{\mathrm{obs}}$ as
\begin{equation}
\begin{aligned}&\Phi_{u,i,t}=\mathrm{LN}\Big(
f_{\mathrm{mod}}\left(\operatorname{Enc}(M_{u,i}),A_{u,i}\right)
\\
&\,\,\,+f_{\mathrm{num}}\left(r_{u,i},\kappa_{u,i}\right)+\phi_i+\phi_t+\phi_{A_{u,i}}
\Big),\end{aligned}
\label{eq:event_encoder}
\end{equation}
where $\mathrm{LN}$ denotes layer normalization. Thus, $\Phi_{u,i,t}$ captures both the observed review content and the modality availability pattern. Appendix~\ref{app:impl:features} specifies $f_{\mathrm{mod}}$ and $f_{\mathrm{num}}$, and Appendix~\ref{app:mnar_gating} isolates the contribution of learned modality gating from direct observation-pattern encoding. %

The encoded representation $\Phi_{u,i,t}$ can vary with time $t$, allowing the relevance of review information to evolve. Appendix~\ref{app:text_dynamic} compares this dynamic encoding with a static alternative.

\newcommand{\causalgraphfigure}{%
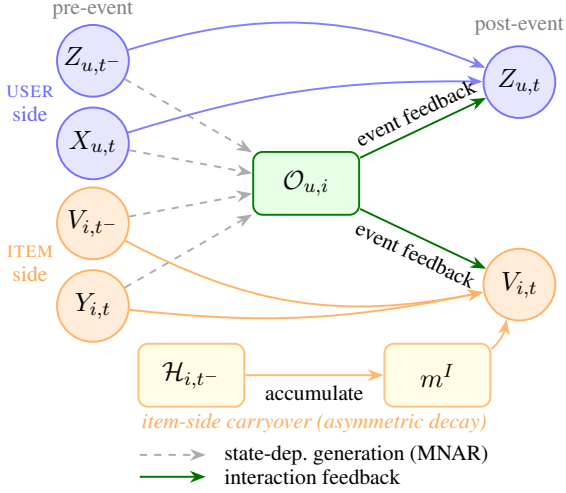
\begin{figure}[t]
\centering
\resizebox{\columnwidth}{!}{%
\begin{tikzpicture}[
    x=0.80cm, y=0.78cm,
    >=Stealth,
    latent/.style={
        circle, draw, thick,
        minimum size=1.05cm, inner sep=1pt,
        font=\normalsize
    },
    obs/.style={
        rectangle, rounded corners=4pt, draw, thick,
        minimum width=1.55cm, minimum height=0.92cm,
        inner sep=4pt, font=\normalsize
    },
    usern/.style={fill=blue!10,   draw=blue!55},
    itemn/.style={fill=orange!15, draw=orange!60},
    obsn/.style ={fill=green!10,  draw=green!50!black},
    memn/.style ={fill=yellow!12, draw=orange!50},
    mnar/.style  ={->, dashed, thick, gray!65},
    uedge/.style ={->, thick, blue!50},
    iedge/.style ={->, thick, orange!60},
    fedge/.style ={->, thick, green!45!black},
    cedge/.style ={->, thick, orange!55}
]

\node[latent, usern] (Zut)  at (-3.9,  2.3) {$Z_{u,t^-}$};
\node[latent, usern] (Xt)   at (-3.9,  0.75){$X_{u,t}$};

\node[latent, itemn] (Vit)  at (-3.9, -0.75){$V_{i,t^-}$};
\node[latent, itemn] (Yt)   at (-3.9, -2.3) {$Y_{i,t}$};

\node[obs, obsn]     (Ouit) at ( 0.0,  0.0) {$\mathcal{O}_{u,i}$};

\node[latent, usern] (Zut1) at ( 3.9,  1.9) {$Z_{u,t}$};
\node[latent, itemn] (Vit1) at ( 3.9, -1.9) {$V_{i,t}$};

\node[obs, memn]     (Hit)  at (-2.1, -3.6) {$\mathcal{H}_{i,t^-}$};
\node[obs, memn]     (Mmem) at ( 2.4, -3.6) {$m^I$};

\draw[mnar] (Zut)  -- (Ouit);
\draw[mnar] (Xt)   -- (Ouit);
\draw[mnar] (Vit)  -- (Ouit);
\draw[mnar] (Yt)   -- (Ouit);

\draw[uedge] (Zut) to[bend left=20]  (Zut1);
\draw[uedge] (Xt)  to[bend left=9]   (Zut1);

\draw[iedge] (Vit) to[bend right=20] (Vit1);
\draw[iedge] (Yt)  to[bend right=9]  (Vit1);

\draw[fedge] (Ouit) -- (Zut1)
    node[midway, above, sloped, font=\footnotesize, black]
        {event feedback};
\draw[fedge] (Ouit) -- (Vit1)
    node[midway, below, sloped, font=\footnotesize, black]
        {event feedback};

\draw[cedge] (Hit)  -- (Mmem)
    node[midway, below, font=\footnotesize, black] {accumulate};
\draw[cedge] (Mmem) to[bend right=22] (Vit1)
    node[near end, right, font=\footnotesize, black] {};

\node[blue!60,   font=\small, align=center] at (-5.05,  1.5)
    {\textsc{user}\\side};
\node[orange!70, font=\small, align=center] at (-5.05, -1.5)
    {\textsc{item}\\side};

\node[gray, font=\small] at (-3.9,  3.25) {pre-event};
\node[gray, font=\small] at ( 3.9,  2.95) {post-event};

\node[orange!65, font=\small, align=center] at ( 0.15, -4.5)
    {\textit{item-side carryover (asymmetric decay)}};

\matrix[
    matrix of nodes,
    ampersand replacement=\&,
    column sep=8pt,
    row sep=2pt,
    nodes={font=\footnotesize, inner sep=0pt, anchor=west, text=black}
] (legend) at (0.15,-5.28) {
    \node[minimum width=0.95cm, minimum height=0.18cm] (legmnar) {}; \& state-dep.\ generation (MNAR) \\
    \node[minimum width=0.95cm, minimum height=0.18cm] (legfb) {}; \& interaction feedback \\
};
\draw[mnar]  (legmnar.west) -- (legmnar.east);
\draw[fedge] (legfb.west) -- (legfb.east);

\end{tikzpicture}
}
\caption{Filtering graph for TS-SSM. \textit{Dashed arrows} encode the predictive dependence of whether a review occurs and, conditional on occurrence, the realized record $\mathcal{O}_{u,i}$ on pre-event states $Z_{u,t^-}$, $V_{i,t^-}$ and contexts $X_{u,t}$, $Y_{i,t}$. \textit{Solid arrows} represent state persistence and feedback from the realized event, encoded by $\Phi_{u,i,t}$, to the post-event states $Z_{u,t}$ and $V_{i,t}$. \textit{Carryover path}: prior item reviews $\mathcal{H}_{i,t^-}$ accumulate into memory $m$ and shape $V_{i,t}$ through asymmetric decay. 
}
\label{fig:causal_graph}
\end{figure}
}

\smallskip
\noindent\textit{Historical-deviation encoding.} The observation pattern and content of a review can also be informative relative to how a user or item has behaved historically. For a current event $e_{u,i}$ at time $t$, we therefore compare the observed review $\mathcal O_{u,i}$ with the pre-event user and item histories, $\mathcal H_{u,t^-}$ and $\mathcal H_{i,t^-}$ to obtain the following outputs:
\begin{equation*}
\begin{aligned}
(\delta^{U}_{u,t},\delta^{I}_{i,t},m_{u,t},r_{u,t})
&=f_{\mathrm{evt}}\big(\mathcal{H}_{u,t^-},\mathcal{H}_{i,t^-},\mathcal O_{u,i}\big).
\end{aligned}
\end{equation*}
Here, $\delta^{U}_{u,t}$ captures deviations from the user's historical expression pattern, while $\delta^{I}_{i,t}$ captures deviations from the item's historical review pattern. The current-event message $m_{u,t}$ summarizes information from $\Phi_{u,i,t}$ used for state updates, and $r_{u,t}$ is a learned reliability-related score based on the current event and its user and item histories. Appendix~\ref{app:impl:update_networks} specifies $f_{\mathrm{evt}}$.

\subsection{User-Side Latent-State Evolution}
\label{sec:method:hsd}
We next specify how TS-SSM learns the user-state evolution in Equation~\eqref{eqn:user-evolution}. For a query event $e_{u,i}$ at time $t$, we construct the post-event user state as 
\begin{equation}\label{eq:user-state-decomp}
    z_{u,t}=z_{u,t^-}+\Delta z^{\mathrm{ng}}_{u,t}+\Delta z^{\mathrm{mp}}_{u,t}.
\end{equation}
Here, $z_{u,t^-}$ summarizes the user's state before the current event; $\Delta z^{\mathrm{ng}}_{u,t}$ captures the non-graph update based on the temporal context $X_{u,t}$, the user's review history $\mathcal H_{u,t^-}$, and the current event $e_{u,i}$; and $\Delta z^{\mathrm{mp}}_{u,t}$ captures the cross-item update propagated through the local user--item graph. We describe each in turn. The transition uses $\mathbf{\Phi}_t$ through the current event, the user's event history, time-aligned item states, and the local graph context.

\subsubsection{Pre-Event User State} 
We encode the user's pre-event review history as
\begin{equation*}
z_{u,t^-} = f^{U}_{\mathrm{enc}}(\mathcal{H}_{u,t^-}).
\end{equation*}
Here, $f^{U}_{\mathrm{enc}}$ is the user-history encoder, which uses representations $\Phi_{u,i',\tau_{u,i'}}$ for events in $\mathcal H_{u,t^-}$. When $\mathcal H_{u,t^-}=\varnothing$, the encoder uses a learned empty-history state. The resulting $z_{u,t^-}$ summarizes user $u$'s review events prior to $e_{u,i}$ and serves as the base state to which information from the current event and other signals at time $t$ is added. Appendix~\ref{app:impl:update_networks} specifies $f^{U}_{\mathrm{enc}}$.

\subsubsection{Within-User State Update} 
We next update $z_{u,t^-}$ using information available from the user's history and the current event. The non-graph update $\Delta z^{\mathrm{ng}}_{u,t}$ combines three sources of information: (i) the current review and the user's prior reviews; (ii) systematic temporal variation; and (iii) the deviation of the current review from the user's historical expression pattern.

\smallskip
\noindent\textit{Current and historical review information.}
For the current review, we use the event-level message $m_{u,t}$ constructed in Section~\ref{sec:method:deviation}. 

For the user's prior reviews, we construct an observation-weighted history-to-user message using a three-step procedure. First, we construct a weight for each prior event $e_{u,i'}\in\mathcal H_{u,t^-}$, 
\begin{equation}
\begin{aligned}
\omega^{U}_{u,i',t}&=w^{U}_{\mathrm{obs},u,i',t}\times w^{U}_{\mathrm{rel},u,i',t}\\
&\quad\times w^{U}_{\mathrm{evid},u,i',t}\times w^{U}_{\mathrm{time},u,i',t},
\end{aligned}
\label{eq:user_msg_weight}
\end{equation}
where the four factors account for the modality-observation pattern, reliability-related cues, feedback magnitude, and recency, respectively. This weighting allows prior reviews to contribute differently depending on both how they were observed and their relevance to the current user state. Appendix~\ref{app:impl:user-weights} provides the corresponding factor specification.

Second, we construct a message for each prior event using the time-aligned item state:
\begin{equation*}
\xi^U_{u,i',t}=W^{U}\big(\Phi_{u,i',\tau_{u,i'}}+v_{i',t}\big),
\end{equation*}
where $W^{U}$ is a learned linear projection, and item state $v_{i',t}$ will be specified in Section~\ref{subsec:item-evolution}. 

Third, we weight the messages of prior events as
\begin{equation}
m^{U}_{u,t}=\frac{\sum_{e_{u,i'}\in\mathcal H_{u,t^-}}\omega^{U}_{u,i',t}\xi^U_{u,i',t}}{\sum_{e_{u,i'}\in\mathcal H_{u,t^-}}\omega^{U}_{u,i',t}}.
\label{eq:user-history-message}
\end{equation}
where the denominator is to normalize the weights $\omega^{U}_{u,i',t}$.
When $\mathcal H_{u,t^-}=\varnothing$, we set $m^U_{u,t}=\mathbf 0$.

\smallskip
\noindent\textit{Systematic temporal variation.}
We encode systematic temporal variation as
\begin{equation}
x_{u,t}=f^{U}_{\mathrm{sys}}\left(\mathcal T^{U}_{u,t}\right) ,
\label{eq:user-temporal-context}
\end{equation}
where $\mathcal T^{U}_{u,t}$ contains observed population- and user-group-level review statistics and their first differences from the time bin immediately preceding $b(t)$. These statistics capture temporal patterns shared across users, and $x_{u,t}$ provides a realization of the systematic temporal context $X_{u,t}$ in Equation~\eqref{eqn:user-evolution}. Appendix~\ref{app:impl:update_networks} specifies $f^{U}_{\mathrm{sys}}$, and Appendix~\ref{app:impl:groups} details the group assignments and lagged temporal contexts.

\smallskip
\noindent\textit{Deviation from historical expression.}
Deviations from a user's historical expression pattern may provide information about changes in the user's state. Using the deviation encoding $\delta^{U}_{u,t}$ from Section~\ref{sec:method:deviation}, we construct a state innovation
\begin{equation*}
h^{U}_{u,t}=f^{U}_{\mathrm{ind}}\left([z_{u,t^-};\,\delta^{U}_{u,t}]\right).
\end{equation*}
where $[\cdot\,;\cdot]$ denotes vector concatenation.
We construct a gate $\gamma^{U}_{u,t}=\sigma(f^{U}_{\mathrm{gate}}([z_{u,t^-};\delta^{U}_{u,t}])),$
where $\sigma$ is the sigmoid function, so that $\gamma^{U}_{u,t}\in[0,1]$. The gate controls how strongly the history-relative innovation $h^{U}_{u,t}$ contributes to the state update. Thus, the effect of the current review on the user state depends on how it differs from the prior expression pattern. Appendix~\ref{app:impl:update_networks} specifies $f^{U}_{\mathrm{ind}}$ and $f^{U}_{\mathrm{gate}}$.

\smallskip
\noindent\textit{Combined within-user update.} We first combine the systematic temporal context and the history-relative innovation as
\begin{equation*}
\Delta z_{u,t}=\eta_2 x_{u,t}+\eta_3\gamma^{U}_{u,t} h^{U}_{u,t},
\end{equation*}
where $\eta_2$ and $\eta_3$ are learned nonnegative scalars. These two components capture the second and third sources of information described above.

We then combine this innovation with the current and historical review information to obtain the complete non-graph update:
\begin{equation}
\begin{aligned}
\Delta z^{\mathrm{ng}}_{u,t}={}&\sigma(r_{u,t})W_{u,t}m_{u,t}+m^U_{u,t}\\
&+\mathrm{Bound}_{\alpha}\left(\Delta z_{u,t};z_{u,t^-}\right),
\end{aligned}
\label{eq:user-nongraph-update}
\end{equation}
where $r_{u,t}$ is the reliability-related score from Section~\ref{sec:method:deviation} and $W_{u,t}$ is a learned projection. The first two terms incorporate the current review and the user's prior reviews, respectively.

The term $\Delta z_{u,t}$ enters Equation~\eqref{eq:user-nongraph-update} through the bounding operator
\begin{equation}
\mathrm{Bound}_{\alpha}(\Delta s; s)=\min\left(1,\frac{\alpha\lVert s\rVert_2}{\max(\lVert\Delta s\rVert_2,\epsilon)}\right)\Delta s.
\label{eq:bounding}
\end{equation}
This operator prevents excessively large changes relative to the accumulated user state while preserving the direction of the proposed update.

\subsubsection{Local Graph Message Passing}
\label{sec:method:encoding} 

A new review may change the user's preferences toward related items. For example, a negative experience with the current item may make a previously reviewed alternative more attractive. We capture relational update through a local user--item graph.

Starting from the pre-propagation state $z^{\mathrm{mid}}_{u,t}=z_{u,t^-}+\Delta z^{\mathrm{ng}}_{u,t}$, we construct a local graph representation $a''_{u,t}$ and propagate it back to the user state:
\begin{equation}
\Delta z^{\mathrm{mp}}_{u,t}=g_{\mathrm{fb}}(z^{\mathrm{mid}}_{u,t},a''_{u,t})W_{\mathrm{fb}}a''_{u,t},
\label{eq:user-mp-update}
\end{equation}
where $g_{\mathrm{fb}}$ controls the strength of this update and $W_{\mathrm{fb}}$ is a learned projection. To construct $a''_{u,t}$, we use the user's most recently reviewed item as an anchor, contextualize it using the current user state, and propagate from the anchor to related items.

\smallskip
\noindent\textit{Anchor-item contextualization.}
When $\mathcal H_{u,t^-}$ is nonempty, let $i^{\mathrm{anc}}_{u,t}$ denote the item in the user's most recent event before time $t$. We use it as a reference for the user's recent preferences, with anchor representation $a_{u,t}=v_{i^{\mathrm{anc}}_{u,t},t}$, where $v_{j,t}$ is the maintained review-conditioned state of item $j$ at time $t$. Section~\ref{subsec:item-evolution} specifies its evolution.

We contextualize the anchor using the within-user state after the current event:
\begin{equation}
a'_{u,t}=a_{u,t}+g_{UI}(z^{\mathrm{mid}}_{u,t},a_{u,t})W^{UI}z^{\mathrm{mid}}_{u,t}.
\label{eq:anchor-update}
\end{equation}
This user-to-item message updates the anchor in light of the user's current preferences: the current review may change how a previously reviewed item serves as a reference point. The gate $g_{UI}$ controls the strength of this update.

\smallskip
\noindent\textit{Propagation to related items.}
The current event may also inform the user's preferences toward items related to the anchor. We therefore augment the contextualized anchor with neighboring item states. Let $\mathcal N(i^{\mathrm{anc}}_{u,t})$ denote the neighbors of the anchor. We compute
\begin{equation}
a''_{u,t}=a'_{u,t}+\sum_{n\in\mathcal N(i^{\mathrm{anc}}_{u,t})}\alpha_n W^{II}v_{n,t},
\label{eq:neighbor-aggregation}
\end{equation}
where $W_{II}$ is a learned projection and
\begin{align*}
\alpha_n&=\operatorname{softmax}_{n\in\mathcal N(i^{\mathrm{anc}}_{u,t})}\left(\frac{(a'_{u,t})^\top v_{n,t}}{\sqrt H}\right).
\end{align*}
Thus, neighbors more aligned with the contextualized anchor receive greater weight. Through Equation~\eqref{eq:user-mp-update}, this local context propagates back to the user state. We maintain at most $K=20$ highest-count neighbors in $\mathcal N(i^{\mathrm{anc}}_{u,t})$ to avoid global graph traversal. More details are provided in Appendix~\ref{app:impl}. 

\subsection{Item-Side Latent-State Evolution}
\label{subsec:item-evolution}
We next specify how TS-SSM learns the item-state evolution in Equation~\eqref{eqn:item-evolution}. For a review event $e_{u,i}$ at time $t$, we construct the post-event item state as
\begin{equation}
v_{i,t}=v_{i,t^-}+\mathrm{Bound}_{\alpha}\left(\Delta v^{I}_{i,t};v_{i,t^-}\right),
\label{eq:item-state-update}
\end{equation}
where $v_{i,t^-}$ summarizes the item's state prior to the current review and $\Delta v^{I}_{i,t}$ captures the item-state update induced by information available at time $t$. The bounding operator is defined in Equation~\eqref{eq:bounding}. 

\subsubsection{Pre-Event Item State}
We encode the pre-event review history as
\begin{equation*}
v_{i,t^-}=f^{I}_{\mathrm{enc}}\left(\mathcal H_{i,t^-}\right).
\end{equation*}
where $f^{I}_{\mathrm{enc}}$ is the item-history encoder and uses a learned empty-history state when $\mathcal H_{i,t^-}=\varnothing$. Appendix~\ref{app:impl:update_networks} specifies $f^{I}_{\mathrm{enc}}$.

\subsubsection{Item-State Update}
The update $\Delta v^{I}_{i,t}$ uses three sources of information: (i) systematic temporal variation; (ii) the deviation of the current review from the item's historical review pattern; and (iii) carryover from prior reviews. 

\smallskip
\noindent\textit{Systematic temporal variation.}
Similar to the user side, we encode item-side temporal variation as
\begin{equation*}
y_{i,t}=f^{I}_{\mathrm{sys}}\left(\mathcal T^{I}_{i,t}\right),
\end{equation*}
where $\mathcal T^{I}_{i,t}$ replaces user-group statistics with item-group statistics, and $y_{i,t}$ realizes the systematic temporal context $Y_{i,t}$ in Equation~\eqref{eqn:item-evolution}. Appendix~\ref{app:impl:update_networks} specifies $f^{I}_{\mathrm{sys}}$, and Appendix~\ref{app:impl:groups} details the group assignments and lagged temporal contexts.

\smallskip
\noindent\textit{Deviation from historical expression.} A current review that deviates from the item's historical review pattern may signal a state change. Using the item-side deviation encoding $\delta^{I}_{i,t}$ from Section~\ref{sec:method:deviation}, we construct a state innovation
\begin{equation*}
h^{I}_{i,t}=f^{I}_{\mathrm{ind}}\left([v_{i,t^-};\,\delta^{I}_{i,t}]\right).
\end{equation*}
We construct a gate $\gamma^{I}_{i,t}=\sigma(f^{I}_{\mathrm{gate}}([v_{i,t^-};\,\delta^{I}_{i,t}])),$
where $\gamma^{I}_{i,t}\in[0,1]$ controls how strongly the history-relative innovation contributes to the item-state update. Thus, the effect of the current review on the item state can depend on how it differs from the item's historical review pattern. Appendix~\ref{app:impl:update_networks} specifies $f^{I}_{\mathrm{ind}}$ and $f^{I}_{\mathrm{gate}}$.

\smallskip
\noindent\textit{Carryover memory.}
Reviews may continue to affect an item's state after the events at which they occur, and positive and negative feedback may persist at different rates. We summarize prior reviews of item $i$ through a carryover memory. Let $t_n<t$ denote the timestamp of prior review $n$. We compute its item-state innovation when the review occurs as
\begin{equation*}
h_n^I=f^{I}_{\mathrm{ind}}\left([v_{i,t_n^-};\,\delta^{I}_{i,t_n}]\right).
\end{equation*}
The innovation $h_n^I$ remains fixed after $t_n$, while its contribution at time $t$ is modulated by the time-dependent weight $\omega^I_{n,t}$. We assign this review the factorized weight
\begin{equation}
\omega^{I}_{n,t}=w^{I}_{\mathrm{obs},n}\times w^{I}_{\mathrm{rel},n,t}\times w^{I}_{\mathrm{sign},n}\times w^{I}_{\mathrm{time},n,t},
\label{eq:carry_weights}
\end{equation}
where the four factors account for the observation pattern, reliability-related cues, feedback magnitude, and recency, respectively. The resulting memory is
\begin{equation}
m^{I}_{i,t}=\frac{\sum_{n:t_n<t}\omega^{I}_{n,t}\tanh(q_n)W^Ih_n^I}{\sum_{n:t_n<t}\omega^{I}_{n,t}},
\label{eq:item-memory}
\end{equation}
where $q_n$ is a learned signed review score and $W^I$ is a learned projection. The recency factor uses separate learned decay rates for positive and negative feedback. Appendices~\ref{app:impl:signed-score} and~\ref{app:impl:carryover} define $q_n$ and detail the weighting factors and decay parameterization, respectively. When $\mathcal H_{i,t^-}=\varnothing$, we set $m^{I}_{i,t}=\mathbf 0$.

\smallskip
\noindent\textit{Combined item update.}
We combine the three components as
\begin{equation}
\Delta v^{I}_{i,t}=\beta_2 y_{i,t}+\beta_3\gamma^{I}_{i,t}h^{I}_{i,t}+\beta_4m^{I}_{i,t},
\label{eq:item-combined-update}
\end{equation}
where $\beta_2$, $\beta_3$, and $\beta_4$ are learned nonnegative scalars. The three terms capture systematic temporal variation, the history-relative innovation from the current review, and persistent effects of prior reviews, respectively. Through Equation~\eqref{eq:item-state-update}, this combined update produces the post-event state $v_{i,t}$.

\causalgraphfigure

\subsection{Item Alignment, Scoring, and Training}
\label{sec:method:training}

Given the final user state $z_{u,t}$, we align non-target candidate items to query time $t$ before scoring. Let $\bar v_{j,t^-}$ denote item $j$'s latest maintained state before $t$. We construct its query-time state as
\begin{equation}
v_{j,t}=\bar v_{j,t^-}+\lambda_{j,t}\Delta v_{j,t},
\label{eq:dynamic_item_state}
\end{equation}
where $\Delta v_{j,t}$ is a temporal adjustment and $\lambda_{j,t}\in[0,1]$ controls its strength:
\begin{align*}
\Delta v_{j,t}&=\operatorname{MLP}_{\mathrm{cand}}\left([\bar v_{j,t^-};\,\phi_j;\,\phi_t]\right),\\
\lambda_{j,t}&=\sigma\left(\operatorname{MLP}_{\mathrm{gate}}\left([\bar v_{j,t^-};\,\Delta v_{j,t}]\right)\right),
\end{align*}
where $\phi_j$ and $\phi_t$ encode item and time features.

For scoring, we use the post-event state $v_{i,t}$ from Section~\ref{subsec:item-evolution} for the target item $i$ and the aligned state $v_{j,t}$ from Equation~\eqref{eq:dynamic_item_state} for each $j\neq i$ in Equation~\eqref{eqn:score}. We decompose its bias term as $\mu_{j,t}=\mu_j+\mu^g_{j,t}$,
where $\mu_j$ is a learned item-specific static bias and $\mu^g_{j,t}$ is a learned dynamic bias shared within item $j$'s group. 

We train TS-SSM using the loss function
\begin{equation*}
\begin{aligned}
\mathcal{L}&=\mathcal{L}_{\mathrm{rank}}+\lambda_{\mathrm{drift}}\mathcal{L}_{\mathrm{drift}}+\lambda_{\mathrm{carry}}\mathcal{L}_{\mathrm{carry}}+\lambda_{\mathrm{rel}}\mathcal{L}_{\mathrm{rel}}.
\end{aligned}
\end{equation*}
The ranking term is temperature-scaled pairwise BPR over valid sampled pairs $\mathcal P$:
{\small \begin{align*}
\mathcal L_{\mathrm{rank}}&=\frac{1}{|\mathcal P|}\sum_{(u,i^+,i^-)\in\mathcal P}\operatorname{softplus}\left(\frac{s(u,i^-)-s(u,i^+)}{\tau}\right).
\end{align*}}

$\mathcal{L}_{\mathrm{drift}}$ and $\mathcal{L}_{\mathrm{carry}}$ predict masked, train-only next-bin item proxies; $\mathcal{L}_{\mathrm{rel}}$ predicts observation, history-support, and user--item conflict proxies. They regularize temporal decomposition alongside the primary ranking loss. Appendices~\ref{app:impl} and~\ref{app:loss_sensitivity} specify targets, masks, schedules, and negative sampling, and evaluate coefficient sensitivity, respectively.

\section{Experiments}
\label{sec:experiments}

\paragraph{Datasets.}
We use six Amazon Reviews 2023 categories~\citep{hou2024bridging}: Toys \& Games, Pet Supplies, Sports \& Outdoors, Electronics, Clothing, and Home \& Kitchen. After validating identifiers, ratings, and timestamps, we remove pre-2014 events and iteratively apply bipartite 10-core filtering, removing users and items with fewer than ten remaining events until convergence. Per-user leave-last-two-out assigns the penultimate and final events to validation and testing, respectively. We use 112 train-aligned time bins: monthly from January 2014 through March 2023, plus one for April--August 2023. We also evaluate on Goodreads Book Graph's Fantasy \& Paranormal subset~\citep{wan2018monotonic,wan2019spoiler}, a non-Amazon platform with different interaction and review distributions. Appendices~\ref{app:datasets} and~\ref{app:goodreads} report the Amazon pipeline and statistics, and the Goodreads results and component ablations, respectively.

\begin{table*}[t]
\centering
\caption{Overall performance on the three primary Amazon datasets. Best results are in \textbf{bold} and second-best results are \underline{underlined}. All methods use the same preprocessing and evaluation protocol.}
\label{tab:main_results}
\small
\setlength{\tabcolsep}{4.5pt}
\resizebox{\textwidth}{!}{%
\begin{tabular}{lcccccccccccc}
\toprule
& \multicolumn{4}{c}{\textbf{Toys \& Games}} & \multicolumn{4}{c}{\textbf{Pet Supplies}} & \multicolumn{4}{c}{\textbf{Sports \& Outdoors}} \\
\cmidrule(lr){2-5}\cmidrule(lr){6-9}\cmidrule(lr){10-13}
Method & R@10 & R@20 & N@10 & N@20 & R@10 & R@20 & N@10 & N@20 & R@10 & R@20 & N@10 & N@20 \\
\midrule
Caser & .0160 & .0267 & .0079 & .0105 & .0165 & .0274 & .0081 & .0108 & .0193 & .0310 & .0097 & .0126 \\
GRU4Rec & .0210 & .0346 & .0105 & .0139 & .0217 & .0357 & .0108 & .0144 & .0311 & .0484 & .0158 & .0201 \\
SASRec & .0649 & .0924 & .0364 & .0433 & .0670 & .0954 & .0376 & .0447 & .0426 & .0646 & .0228 & .0283 \\
BERT4Rec & .0632 & .0934 & .0351 & .0428 & .0653 & .0965 & .0363 & .0442 & .0203 & .0318 & .0109 & .0139 \\
FMLPRec & .0667 & .0969 & .0363 & .0439 & .0689 & .1001 & .0375 & .0453 & .0190 & .0297 & .0099 & .0125 \\
VBPR & .0456 & .0717 & .0240 & .0304 & .0478 & .0752 & .0252 & .0318 & .0287 & .0449 & .0155 & .0196 \\
MMGCN & .0570 & .0838 & .0311 & .0376 & .0598 & .0879 & .0325 & .0394 & .0348 & .0535 & .0191 & .0240 \\
LATTICE & .0732 & .1019 & .0416 & .0487 & .0768 & .1069 & .0436 & .0511 & .0362 & .0553 & .0203 & .0254 \\
BM3 & .0764 & .1062 & .0434 & .0508 & .0801 & .1114 & .0454 & .0532 & .0359 & .0545 & .0197 & .0246 \\
CL4SRec & .0889 & .1197 & .0518 & .0598 & .0926 & .1246 & .0540 & .0623 & .0473 & .0688 & .0281 & .0339 \\
DuoRec & .1027 & .1362 & .0650 & .0733 & .1071 & .1419 & .0676 & .0764 & .0571 & .0808 & .0370 & .0436 \\
FEARec & .1047 & .1390 & .0658 & .0743 & .1096 & .1452 & .0687 & .0777 & .0573 & .0820 & .0363 & .0430 \\
LLMRec & .1023 & .1351 & .0642 & .0725 & .1073 & .1419 & .0673 & .0761 & .0539 & .0764 & .0329 & .0392 \\
POD & .1056 & .1401 & .0665 & .0750 & .1109 & .1470 & .0697 & .0787 & .0593 & .0831 & .0378 & .0443 \\
SAID & .1065 & .1414 & .0670 & .0757 & .1119 & .1484 & .0703 & .0795 & .0573 & .0817 & .0417 & .0475 \\
BSARec & .1063 & .1427 & .0746 & .0838 & .1118 & .1500 & .0784 & .0880 & .0624 & .0838 & .0438 & .0493 \\
RecGPT & .1080 & .1463 & .0752 & .0848 & .1127 & .1527 & .0785 & .0885 & .0634 & .0859 & .0442 & .0500 \\
HORAE & .1086 & .1471 & .0757 & .0854 & .1139 & .1544 & .0794 & .0896 & .0638 & .0864 & .0445 & .0504 \\
HM4SR & \underline{.1099} & \underline{.1489} & \underline{.0766} & \underline{.0865} & \underline{.1153} & \underline{.1563} & \underline{.0804} & \underline{.0908} & \underline{.0644} & \underline{.0872} & \underline{.0448} & \underline{.0507} \\
\midrule
\textbf{TS-SSM} & \textbf{.1162} & \textbf{.1683} & \textbf{.0781} & \textbf{.0912} & \textbf{.1212} & \textbf{.1756} & \textbf{.0815} & \textbf{.0953} & \textbf{.0664} & \textbf{.0962} & \textbf{.0453} & \textbf{.0530} \\
\bottomrule
\end{tabular}
}
\end{table*}

\paragraph{Baselines.}
\label{sec:exp:baseline}
On the three primary Amazon datasets, we compare four families:
\textit{Sequential}: Caser~\citep{tang2018personalized}, GRU4Rec~\citep{hidasi2016session}, SASRec~\citep{kang2018self}, BERT4Rec~\citep{sun2019bert4rec}, FMLPRec~\citep{zhou2022filter}, CL4SRec~\citep{xie2022contrastive}, DuoRec~\citep{qiu2022contrastive}, FEARec~\citep{du2024frequency}, and BSARec~\citep{shin2024bsarec};
\textit{Multimodal}: VBPR~\citep{he2016vbpr}, MMGCN~\citep{wei2019mmgcn}, LATTICE~\citep{zhang2021mining}, BM3~\citep{zhou2023bootstrap}, and LLMRec~\citep{wei2024llmrec};
\textit{Adjustment and auxiliary signal}: POD~\citep{li2023prompt} and SAID~\citep{hu2024said}; and
\textit{Recent sequential and multimodal}: RecGPT~\citep{jiang2025recgpt}, HORAE~\citep{hu2025horae}, and HM4SR~\citep{zhang2025hm4sr}.
BSARec is the primary established sequential reference, while HM4SR has the highest average Recall@20 among recent baselines. We also augment BSARec with a conditioning token comprising TS-SSM's modality-availability pattern, rating, and time bin. This \textit{BSARec + current-event features} control tests how much these three metadata fields explain the gains (Appendix~\ref{app:info_boundary}). On the other three Amazon datasets, we evaluate BSARec, MACR~\citep{wei2021model}, AutoDebias~\citep{chen2021autodebias}, RecGPT, HORAE, and HM4SR.

\paragraph{Evaluation protocol.}
Following standard practice~\citep{kang2018self,sun2019bert4rec}, we use full-item ranking and validation Recall@20 checkpoint selection. We report Recall@10/20 and NDCG@10/20, with Recall/NDCG@50 and MRR in extended results. Because published Amazon-review evaluations differ in core filtering, rating thresholds, date ranges, temporal discretization, and negative sampling, and no standardized leaderboard covers them, we reproduce every baseline using TS-SSM's iterative bipartite 10-core filtering, per-user leave-last-two-out split, January-2014 cutoff, and 112-bin construction. Unless noted, full-sort filters items the user observed before the target timestamp while retaining and separately scoring the ground-truth item. Appendices~\ref{app:impl} and~\ref{app:significance} provide exact rank computation and multi-run paired $t$-tests with Bonferroni-adjusted decisions, confidence intervals, and effect sizes, respectively.

\subsection{Main Results}
\label{sec:exp:main}

Table~\ref{tab:main_results} compares TS-SSM with all baselines on three primary Amazon datasets. TS-SSM is best on all twelve metrics, achieving Recall@20 of .1683, .1756, and .0962 on Toys, Pets, and Sports, respectively—relative gains of $17.9\%$, $17.1\%$, and $14.8\%$ over BSARec. It also outperforms the strongest recent baseline, HM4SR, whose scores are .1489, .1563, and .0872, by $13.0\%$, $12.3\%$, and $10.3\%$. Appendix~\ref{app:info_boundary} shows that adding modality availability, rating, and time-bin information to BSARec improves it by at most $4.7\%$, still leaving a $9.0\%$--$11.3\%$ gap to TS-SSM.

Static multimodal baselines VBPR, MMGCN, and BM3 trail the strongest sequential methods. Although RecGPT, HORAE, and HM4SR progressively improve over BSARec, their gains under the same protocol remain smaller than those of TS-SSM. BM3 has a comparable parameter count, while several recent multimodal systems have higher measured runtime than TS-SSM (Appendix~\ref{app:efficiency}). Together with the BSARec current-event control, these comparisons place the effects of model scale, review-side inputs, and temporal architecture in a common empirical context.

\paragraph{Extended evaluations.}
Table~\ref{tab:cross_dataset} summarizes Recall@20 across six Amazon categories, each with a distinct user--item graph, product vocabulary, and multimodal expression distribution; image-upload rates range from $5.3\%$ to $18.7\%$. TS-SSM outperforms BSARec on every dataset by $14.8\%$--$18.8\%$ and HM4SR by $11.7\%$ on average. Clothing has both the highest image rate ($18.7\%$) and largest BSARec gain ($18.8\%$). Image rate and relative gain correlate at $r=0.67$ across categories (Appendix~\ref{app:richness}), but with only six category-level observations, this is descriptive rather than evidence of a general relationship. Appendices~\ref{app:additional_datasets} and~\ref{app:significance} provide the additional-category metric breakdown and multi-run robustness, respectively.

\begin{table}[t]
\centering
\caption{Amazon R@20 (6-category common setup).}
\label{tab:cross_dataset}
\small
\resizebox{\columnwidth}{!}{%
\begin{tabular}{lccccc}
\toprule
Dataset & BSARec & RecGPT & HORAE & HM4SR & TS-SSM \\
\midrule
Toys \& Games & .1427 & .1463 & .1471 & \underline{.1489} & \textbf{.1683} \\
Pet Supplies & .1500 & .1527 & .1544 & \underline{.1563} & \textbf{.1756} \\
Sports \& Outdoors & .0838 & .0859 & .0864 & \underline{.0872} & \textbf{.0962} \\
Electronics & .1317 & .1342 & .1358 & \underline{.1375} & \textbf{.1530} \\
Clothing & .1391 & .1413 & .1443 & \underline{.1461} & \textbf{.1653} \\
Home \& Kitchen & .1365 & .1390 & .1408 & \underline{.1425} & \textbf{.1570} \\
\bottomrule
\end{tabular}
}
\end{table}

Goodreads Fantasy mirrors Amazon: TS-SSM reaches .5847 Recall@20 versus HM4SR's .5191, a $12.6\%$ relative gain. Removing item-state updating causes the largest drop ($6.81\%$), followed by message passing ($4.70\%$), preserving the Amazon component ordering. Appendix~\ref{app:goodreads} reports all four metrics and the full ablation.

\begin{table}[t]
\centering
\caption{Toys \& Games ($\Delta$R@20 vs.\ full).}
\label{tab:ablation}
\small
\setlength{\tabcolsep}{5pt}
\resizebox{\columnwidth}{!}{%
\begin{tabular}{lccccc}
\toprule
Variant & R@10 & R@20 & N@10 & N@20 & $\Delta$R@20 \\
\midrule
\textbf{TS-SSM (Full)} & \textbf{.1162} & \textbf{.1683} & \textbf{.0781} & \textbf{.0912} & --- \\
\multicolumn{6}{l}{\textit{User-side components}} \\
w/o User State Update & .1133 & .1641 & .0759 & .0887 & $-2.50\%$ \\
w/o Deviation Features & .1140 & .1651 & .0765 & .0894 & $-1.90\%$ \\
\midrule
\multicolumn{6}{l}{\textit{Item-side components}} \\
w/o Item State Update & .1119 & .1624 & .0751 & .0878 & $-3.50\%$ \\
w/o Message Passing & .1130 & .1636 & .0758 & .0885 & $-2.79\%$ \\
w/o Carryover Memory & .1137 & .1646 & .0764 & .0892 & $-2.20\%$ \\
\midrule
\multicolumn{6}{l}{\textit{Auxiliary supervision}} \\
w/o Auxiliary Losses & .1147 & .1661 & .0771 & .0900 & $-1.31\%$ \\
\bottomrule
\end{tabular}
}
\end{table}

\subsection{Ablation Study}
\label{sec:exp:ablation}

Table~\ref{tab:ablation} decomposes TS-SSM on Toys \& Games by removing one modeling component at a time.
We report the $\Delta$R@20 drop relative to the full model.

Amazon Recall@20 drops most without \textit{Item State Update} ($-3.50\%$), followed by \textit{Message Passing} ($-2.79\%$) and \textit{User State Update} ($-2.50\%$). These components respectively track review histories, inject prefix and related-item signals, and distinguish the current event from the user's expression pattern. Removing \textit{Carryover Memory}, \textit{Deviation Features}, or \textit{Auxiliary Losses} reduces performance by $2.20\%$, $1.90\%$, and $1.31\%$, supporting complementary roles for persistent feedback, MNAR-related deviation cues, and temporal regularization. 
Appendix~\ref{app:ablation_extended} examines grouping, gate, behavioral-deviation, and symmetric-carryover ablations. Over tenfold changes to each auxiliary coefficient reduce Recall@20 by at most $1.43\%$ (Appendix~\ref{app:loss_sensitivity}). Replacing MNAR gating with indicators lowers Recall@20 by $2.97\%$ ($p<10^{-7}$); no-pattern remains $13.0\%$ above BSARec (Appendix~\ref{app:mnar_gating}). Static item-level and no-text variants lower Recall@20 by $2.67\%$ and $6.30\%$ (Appendix~\ref{app:text_dynamic}); larger text/image encoders gain only $2.14\%$ and $0.71\%$--$1.27\%$ (Appendix~\ref{app:encoder_sensitivity}).

\section{Related Work}
\label{sec:related}

Sequential recommendation spans recurrent and convolutional models~\citep{hidasi2016session,tang2018personalized}, Transformers~\citep{kang2018self,sun2019bert4rec}, and foundation-model extensions~\citep{jiang2025recgpt,na2024enhancing,lyu2024llmrec}. Multimodal methods exploit modality \emph{content}~\citep{he2016vbpr,wei2019mmgcn}; missing-modality methods impute or marginalize absent inputs~\citep{guo2024multimodal}. Clinical multimodal models use non-random modality availability for representation~\citep{liang-etal-2025-causal,liang-etal-2026-learning}. TS-SSM transfers this view to behavior-dependent modality use and expression in a two-sided platform-review state space.

Recent baselines are RecGPT, HORAE, and HM4SR~\citep{jiang2025recgpt,hu2025horae,zhang2025hm4sr}. Unlike TPNE's time-dependent excitation~\citep{huang2023tpne}, TS-SSM retains signed review shocks in item-side memory with independently learned decay. Causal recommenders address logged-feedback biases~\citep{schnabel2016recommendations,wang2019doubly,bonner2018causal,wei2021model,wang2021deconfounded}; our predictive correction instead resembles hierarchical decomposition~\citep{gelman2006data}. Appendix~\ref{app:related} provides the extended comparison.

\section{Conclusion}
\label{sec:conclusion}
TS-SSM combines dynamic event-level review encoding, observation-pattern-aware local propagation, joint user--item updates, and asymmetric carryover memory in a two-sided state-space framework. Recall@20 improves by $14.8\%$--$18.8\%$ over BSARec across six Amazon categories and by $11.7\%$ on average over HM4SR ($12.6\%$ on Goodreads Fantasy). Sensitivity and ablations support distinct component roles.

\section{Limitations}
\label{limitation}
We evaluate TS-SSM across six Amazon categories and one non-Amazon Goodreads domain. Broader validation is needed on platforms with sparser or qualitatively different review behavior. Its advantage may diminish when text is uninformative, multimodal expression is rare, or histories are too short to estimate user-relative deviations reliably.

The MNAR formulation is predictive rather than causally identified. Bin-level population contexts represent shared temporal shifts, while user-specific time constraints, interface exposure differences, and other unobserved factors may also shape review absence. The learned observation weights are therefore predictive quantities rather than causal propensities.

TS-SSM requires more computation than lightweight user-only sequential models such as SASRec and BSARec. Maintaining catalog-wide dynamic item states and carryover summaries adds storage and update costs, while bounded local propagation and ANN retrieval can limit online ranking work. These costs depend on the serving environment. The item--item graph uses fixed-window co-occurrence, and the auxiliary targets for drift, carryover, and reliability are predictive proxies rather than externally validated latent variables.

\section*{Acknowledgment}

This research was supported by the URC Award at Emory University.

\bibliography{custom}

\clearpage
\appendix

\section*{Use of Artificial Intelligence}
\label{app:ai_use}

The authors used large language model (LLM)-based writing assistants to help polish the prose of this paper. All technical content, experimental design, results, and conclusions are entirely the work of the authors. The LLM tools were used solely for language editing and were not involved in any aspect of the research, methodology, or data analysis.

\section*{Appendix Overview}
\label{app:toc}

\begin{itemize}[leftmargin=*, itemsep=2pt]
    \item[\ref{app:datasets}] \textbf{Dataset Details} -- Summary statistics for all six Amazon categories and the common evaluation protocol.
    \item[\ref{app:impl}] \textbf{Implementation Details and Hyperparameters} -- TS-SSM settings, multimodal feature extraction, and baseline configurations.
    \item[\ref{app:deviation_analysis}] \textbf{Behavioral Deviation Analysis} -- Empirical motivation for deviation-aware modeling from review-expression patterns.
    \item[\ref{app:additional_datasets}] \textbf{Additional Amazon Categories} -- Compact comparisons on three additional categories.
    \item[\ref{app:goodreads}] \textbf{Cross-Platform Evaluation} -- Results and component ablations on Goodreads Fantasy.
    \item[\ref{app:richness}] \textbf{Review Richness and Cross-Category Gain} -- Descriptive analysis across the six Amazon categories.
    \item[\ref{app:significance}] \textbf{Statistical Robustness} -- Multi-run significance tests on Toys \& Games and Pet Supplies.
    \item[\ref{app:info_boundary}] \textbf{Information Boundary Controls} -- Controls for current-event conditioning and item-identity leakage.
    \item[\ref{app:efficiency}] \textbf{Model Scale, Efficiency, and Deployment} -- Parameter counts, training cost, inference latency, and practical serving considerations.
    \item[\ref{app:ablation_extended}] \textbf{Extended Ablation Study} -- Grouping alternatives, gate variants, deviation signals, and symmetric carryover.
    \item[\ref{app:sensitivity}] \textbf{Sensitivity Analysis} -- Hyperparameter sensitivity for the main design choices.
    \item[\ref{app:loss_sensitivity}] \textbf{Auxiliary-Loss Weight Sensitivity} -- One-at-a-time sweeps for drift, carryover, and reliability coefficients.
    \item[\ref{app:encoder_sensitivity}] \textbf{Encoder Backbone Sensitivity} -- Text- and image-encoder capacity comparisons.
    \item[\ref{app:mnar_gating}] \textbf{MNAR-Gating versus Simple Observation-Pattern Features} -- Comparison against simpler missingness-feature baselines.
    \item[\ref{app:text_dynamic}] \textbf{Review Text as a Dynamic versus Static Signal} -- Event-level text against static item-level text representations.
    \item[\ref{app:subgroup}] \textbf{Subgroup Analysis} -- Reliability-gate patterns and performance by observable history length.
    \item[\ref{app:longtail}] \textbf{Long-Tail Item Analysis} -- Popularity-stratified performance on Toys \& Games.
    \item[\ref{app:case_study}] \textbf{Qualitative Analysis: Gate Activation and Carryover Persistence} -- Representative user-side and item-side case studies.
    \item[\ref{app:error_analysis}] \textbf{Error Analysis} -- Taxonomy of the model's dominant prediction failures.
    \item[\ref{app:related}] \textbf{Extended Related Work} -- Additional discussion of sequential, multimodal, MNAR, temporal, causal, and LLM-based recommendation.
\end{itemize}

\vspace{1em}
\hrule
\vspace{1em}
\newpage
\raggedbottom

\section{Dataset Details}
\label{app:datasets}

All six Amazon categories use the same preprocessing order: category-specific review loading; required-field and timestamp validation; removal of events before 2014; iterative bipartite 10-core filtering; contiguous user/item indexing; chronological history construction; per-user leave-last-two-out; and construction of train-only statistics, groups, temporal summaries, item histories, item neighbors, and auxiliary targets. The 10-core operation repeatedly removes users and items with fewer than ten events in the current bipartite graph until both sides stabilize. Modality extraction and availability encoding do not change split membership.

For Toys \& Games, we use the \texttt{raw\_review\_Toys\_and\_Games} category from \texttt{McAuley-Lab/Amazon-Reviews-2023}. A record is retained only if it has valid user and item identifiers, rating, and timestamp. Timestamps are detected as seconds or milliseconds and converted to datetimes; unparsable values, dates before 1995, dates after preprocessing time, and events before 2014 are removed. User and initial item histories are sorted by Unix time with the stable source \texttt{global\_idx} breaking exact ties. Query-time item histories use only events with timestamps strictly earlier than the target, excluding simultaneous item reviews.

Review title, body, and image are optional and do not determine event retention. Empty title/body fields set \texttt{has\_title=0} or \texttt{has\_text=0} and use zero embeddings; missing images similarly set \texttt{has\_image=0}. The three availability bits form an explicit modality-pattern embedding. After filtering, each user sequence is ordered chronologically: the last event is the test target, the penultimate event is the validation target, and all earlier events form the training split. The test user prefix includes the earlier validation event because it is observable at test query time.

\begin{table}[ht]
\centering
\caption{Summary statistics for all six Amazon categories used in the experiments.}
\label{tab:dataset_all6}
\small
\resizebox{\columnwidth}{!}{%
\begin{tabular}{lccccc}
\toprule
Dataset & \#Users & \#Items & \#Inter. & Img Rate & Sparsity \\
\midrule
Toys \& Games & 35,690 & 28,530 & 611,908 & 8.1\% & 99.94\% \\
Pet Supplies & 38,427 & 31,364 & 674,821 & 10.2\% & 99.94\% \\
Sports \& Outdoors & 27,832 & 35,284 & 473,144 & 5.3\% & 99.95\% \\
Electronics & 42,831 & 35,672 & 724,519 & 12.4\% & 99.95\% \\
Clothing & 38,294 & 31,458 & 583,621 & 18.7\% & 99.95\% \\
Home \& Kitchen & 45,162 & 38,941 & 812,374 & 9.8\% & 99.95\% \\
\bottomrule
\end{tabular}
}
\end{table}

\begin{table*}[ht]
\centering
\caption{Detailed statistics for the three primary Amazon datasets.}
\label{tab:dataset_detail}
\small
\begin{tabular}{lccc}
\toprule
Statistic & Toys & Pets & Sports \\
\midrule
Users $|\mathcal{U}|$ & 35,690 & 38,427 & 27,832 \\
Items $|\mathcal{I}|$ & 28,530 & 31,364 & 35,284 \\
Interactions & 611,908 & 674,821 & 473,144 \\
Sparsity & 99.94\% & 99.94\% & 99.95\% \\
Time span & Jan 2014--Aug 2023 & Jan 2014--Aug 2023 & Jan 2014--Aug 2023 \\
Time bins $T$ & 112 & 112 & 112 \\
Avg.\ interactions per user & 17.1 & 17.6 & 17.0 \\
Avg.\ interactions per item & 21.4 & 21.5 & 13.4 \\
Has title rate & 100.00\% & 100.00\% & 100.00\% \\
Has text rate & 99.97\% & 99.96\% & 99.97\% \\
Has image rate & 8.1\% & 10.2\% & 5.3\% \\
Train / Val / Test & 540,528 / 35,690 / 35,690 & 597,967 / 38,427 / 38,427 & 417,480 / 27,832 / 27,832 \\
\bottomrule
\end{tabular}
\end{table*}

\section{Implementation Details and Hyperparameters}
\label{app:impl}

\subsection{TS-SSM Configuration}
\label{app:impl:hsdmpc}

Table~\ref{tab:hyperparams} lists the TS-SSM settings used in all reported experiments. Training uses AdamW, cosine scheduling, mixed negative sampling, and full-sort checkpoint selection by validation Recall@20. Temporal statistics, group assignments, item graphs, and auxiliary targets are computed from the training split. Query histories obey chronological availability: a validation query cannot use itself or later events, while the earlier validation event is observable in the prefix of the subsequent test query.

\begin{table}[ht]
\centering
\caption{TS-SSM hyperparameter settings.}
\label{tab:hyperparams}
\small
\resizebox{\columnwidth}{!}{%
\begin{tabular}{lc}
\toprule
Hyperparameter & Value \\
\midrule
Hidden dimension $H$ & 320 \\
User Transformer layers & 3 \\
Item Transformer layers & 2 \\
Attention heads & 4 \\
Dropout rate & 0.1 \\
Maximum user sequence length & 64 \\
Maximum item sequence length & 40 \\
Batch size & 128 \\
Learning rate & $5{\times}10^{-4}$ \\
Weight decay & $1{\times}10^{-4}$ \\
Optimizer & AdamW \\
LR scheduler & Cosine annealing \\
Training epochs & 50 \\
Gradient clipping & 1.0 \\
Training negatives & 48 (mixed sampling) \\
Uniform negative ratio & 0.6 \\
Ranking loss weight $\lambda_\text{rank}$ & 1.0 \\
Drift loss weight $\lambda_\text{drift}$ & 0.08 \\
Carry loss weight $\lambda_\text{carry}$ & 0.05 \\
Reliability loss weight $\lambda_\text{rel}$ & 0.03 \\
User groups $G$ & 3 \\
Item groups $G^I$ & 8 \\
Shrinkage $\lambda_s$ & 50 \\
Initial / final norm bound $\alpha$ & 0.10 / 0.15 \\
Warmup epochs & 8 \\
BPR temperature schedule & $1.0 \to 0.5$ \\
Carryover $\lambda_+$ init. & 0.25 \\
Carryover $\lambda_-$ init. & 0.15 \\
User-history $\lambda_U$ init. & 0.20 \\
Item neighbors $K$ & 20 (co-occurrence window 3) \\
EMA decay & 0.999 \\
\bottomrule
\end{tabular}
}
\end{table}

\subsection{Abstract-to-Implementation Correspondence}
\label{app:impl:correspondence}

The collection $\mathbf{\Phi}_t$ in Equations~\eqref{eqn:user-evolution}--\eqref{eqn:item-evolution} contains the event representations available by time $t$. Reviews are processed in chronological order. Each event $e_{u,i}$ at time $t=\tau_{u,i}$ uses lagged statistics from bin $b(t)-1$ for the systematic contexts, constructs the current user-side correction, and updates the maintained state of item $i$. The first bin uses default training statistics. The post-event item state is stored and carried into later events, so the collection of item states records the review history accumulated for every item.

\paragraph{Event-level signed review score.}
\label{app:impl:signed-score}
For an event $e=(u_e,i_e)$ occurring at time $t_e$, let $\widetilde{\Phi}_e$ denote its event representation with the item-identity channel removed. We compute a signed review score once, when the event occurs:
\begin{align*}
q_e={}&\rho_r\bar r_e+a_q^\top\operatorname{GELU}\Big(
W_q[\widetilde{\Phi}_e;\delta^U_{u_e,t_e};\\
&\hspace{39mm}\delta^I_{i_e,t_e}]+b_q\Big)+c_q,\\
\rho_r&=\operatorname{softplus}(\rho_r^{\mathrm{raw}}),\qquad \bar r_e=(r_e-3)/2.
\end{align*}
Here $W_q$, $a_q$, $b_q$, and $c_q$ are learned parameters. The positive rating coefficient anchors the score to rating polarity, while the learned residual incorporates multimodal content and the event's deviations from the user and item histories. The score is stored with the event and remains fixed; only its time-dependent weight changes for later queries.

\paragraph{User-history weighting.}
\label{app:impl:user-weights}
For a query event $e_{u,i}$ at time $t$, define the query context
\begin{equation*}
C_{u,i,t}=[\delta^U_{u,t};\delta^I_{i,t};\chi^U_{t^-};\chi^I_{t^-}].
\end{equation*}
For each prior event $e=e_{u,i'}\in\mathcal H_{u,t^-}$, let $c^U_{e,t}=[x^{\mathrm{num}}_e;C_{u,i,t}]$ and $\Delta b_{e,t}=b(t)-b(t_e)$. Using $e$ as a compact index, its four weighting factors are
\begin{align*}
w^U_{\mathrm{obs},e}&=\epsilon_w+\operatorname{softplus}\left((a^U_{\mathrm{obs}})^\top\phi_{A_e}+b^U_{\mathrm{obs}}\right),\\
w^U_{\mathrm{rel},e,t}&=\epsilon_w+(1-\epsilon_w)\sigma\left(f^U_{\mathrm{rel}}(c^U_{e,t})\right),\\
w^U_{\mathrm{evid},e}&=1+|q_e|,\\
w^U_{\mathrm{time},e,t}&=\exp\left(-\lambda_U\Delta b_{e,t}\right),\\
\lambda_U&=\operatorname{softplus}(\lambda_U^{\mathrm{raw}}),
\end{align*}
where $0<\epsilon_w<1$ is a fixed positive floor, $q_e$ is the stored event-level score in Appendix~\ref{app:impl:signed-score}, and $f^U_{\mathrm{rel}}$ is a scalar MLP. The decay rate $\lambda_U$ is initialized to $0.20$ (Table~\ref{tab:hyperparams}) and learned thereafter. These are the factors indexed by $(u,i',t)$ in Equation~\eqref{eq:user_msg_weight}; the observation and evidence factors depend only on the stored event and can be cached. The observation factor assigns a separate learned weight to each of the eight modality-availability patterns. The reliability factor combines numeric features of the prior event with current query-level deviations and history support. It is distinct from $r_{u,t}$, which scales the current-event message in Equation~\eqref{eq:user-nongraph-update}. The evidence factor changes only the magnitude of the user-history contribution; rating and review-content channels in the event message retain its directional information. The user-side decay is shared across feedback signs, leaving sign-specific persistence to the item-side carryover.

Write $\xi^U_e=\xi^U_{u,i',t}$ and $\omega^U_e=\omega^U_{u,i',t}$, where
\begin{equation*}
\xi^U_{u,i',t}=W^U(\Phi_{u,i',\tau_{u,i'}}+v_{i',t}).
\end{equation*}
The normalized, observation-pattern-weighted user-history message is
\begin{equation*}
m^U_{u,t}=\frac{\sum_{e\in\mathcal H_{u,t^-}}\omega^U_e\xi^U_e}{\sum_{e\in\mathcal H_{u,t^-}}\omega^U_e},
\end{equation*}
where $\omega^U_e=w^U_{\mathrm{obs},e}w^U_{\mathrm{rel},e,t}w^U_{\mathrm{evid},e}w^U_{\mathrm{time},e,t}$, equivalently $\omega^U_{u,i',t}$ in Equation~\eqref{eq:user_msg_weight}. All four factors are positive. The current-event correction and bounded local graph context provide the other user-side pathways from $\mathbf{\Phi}_t$.

On the item side, let $t_n<t$ denote the timestamp of prior review $n$ of item $i$, and define $\xi^I_n=\tanh(q_n)W^Ih_n^I$. The signed carryover summary is
\begin{equation*}
m^I_{i,t}=\frac{\sum_{n:t_n<t}\omega^I_{n,t}\xi^I_n}{\sum_{n:t_n<t}\omega^I_{n,t}},
\end{equation*}
where $\omega^I_{n,t}=w^I_{\mathrm{obs},n}w^I_{\mathrm{rel},n,t}w^I_{\mathrm{sign},n}w^I_{\mathrm{time},n,t}$. The factors and the sign-specific temporal parameterization are detailed below. When the corresponding history is empty, the normalized aggregation is omitted and the memory is set to zero.

The abstract user filtering map $f^U$ is the composition of the bounded state update, weighted history aggregation, current-event correction, and local feedback:
\begin{align*}
\Delta z_{u,t}&=\eta_2x_{u,t}+\eta_3\gamma^U_{u,t}h^U_{u,t},\\
\Delta z^{\mathrm{ng}}_{u,t}&=\sigma(r_{u,t})W_{u,t}m_{u,t}+m^U_{u,t}\\
&\quad+\mathrm{Bound}_\alpha(\Delta z_{u,t};z_{u,t^-}),\\
z^{\mathrm{mid}}_{u,t}&=z_{u,t^-}+\Delta z^{\mathrm{ng}}_{u,t},\\
\Delta z^{\mathrm{mp}}_{u,t}&=g_{\mathrm{fb}}(z^{\mathrm{mid}}_{u,t},a''_{u,t})W_{\mathrm{fb}}a''_{u,t},\\
f^U_{\mathrm{impl}}&=z_{u,t}=z_{u,t^-}+\Delta z^{\mathrm{ng}}_{u,t}+\Delta z^{\mathrm{mp}}_{u,t}.
\end{align*}
Here $z_{u,t^-}$ represents the pre-event state $Z_{u,t^-}$, $x_{u,t}$ realizes the user-specific temporal context $X_{u,t}$ from lagged bin statistics, and the remaining terms select and aggregate the relevant event information from $\mathbf{\Phi}_t$. The final $z_{u,t}$ represents the post-event filtered state $Z_{u,t}$. Similarly, the abstract item filtering map is instantiated as
\begin{align*}
\Delta v^I_{i,t}&=\beta_2 y_{i,t}+\beta_3\gamma^I_{i,t}h^I_{i,t}+\beta_4m^I_{i,t},\\
f^I_{\mathrm{impl}}&=v_{i,t}=v_{i,t^-}+\mathrm{Bound}_\alpha(\Delta v^I_{i,t};v_{i,t^-}),
\end{align*}
where $v_{i,t^-}$ represents the maintained pre-event state $V_{i,t^-}$, $y_{i,t}$ realizes $Y_{i,t}$ from lagged bin statistics, and the gated innovation and carryover memory aggregate the current and prior event information for item $i$ from $\mathbf{\Phi}_t$. The result $f^I_{\mathrm{impl}}=v_{i,t}$ is stored as $V_{i,t}$ and used directly for target scoring. Repeated application of this update maintains the dynamic states of all items. The local propagation in Section~\ref{sec:method:encoding} refines the user branch using prefix and neighboring item states without re-processing the current target state. The process-noise terms $\varepsilon^U$ and $\varepsilon^I$ express residual uncertainty at the abstract-model level.

\paragraph{Carryover-memory parameterization.}
\label{app:impl:carryover}
For prior review $n$ of item $i$ occurring at time $t_n<t$, the associated historical innovation is computed when that review occurs by the same item-side innovation map used for the current review:
\begin{equation*}
h_n^I=f^I_{\mathrm{ind}}\left([v_{i,t_n^-};\,\delta^I_{i,t_n}]\right).
\end{equation*}
The innovation $h_n^I$ remains fixed after $t_n$; its contribution to the carryover memory at time $t$ changes through $\omega^I_{n,t}$. Its carryover weight is
\begin{equation*}
\omega^I_{n,t}=w^I_{\mathrm{obs},n}w^I_{\mathrm{rel},n,t}w^I_{\mathrm{sign},n}w^I_{\mathrm{time},n,t},
\end{equation*}
For the current query event $e_{u,i}$, write $x^{\mathrm{num}}_n$ and $A_n$ for review $n$'s numeric features and availability pattern, and let $c^I_{n,t}=[x^{\mathrm{num}}_n;C_{u,i,t}]$. The four factors are
\begin{align*}
w^I_{\mathrm{obs},n}&=\epsilon_w+\operatorname{softplus}\left((a^I_{\mathrm{obs}})^\top\phi_{A_n}+b^I_{\mathrm{obs}}\right),\\
w^I_{\mathrm{rel},n,t}&=\epsilon_w+(1-\epsilon_w)\sigma\left(f^I_{\mathrm{rel}}(c^I_{n,t})\right),\\
w^I_{\mathrm{sign},n}&=1+|q_n|,\\
w^I_{\mathrm{time},n,t}&=\exp\left[-\lambda_{\operatorname{sign}(q_n)}\Delta b_{n,t}\right],
\end{align*}
where $q_n=q_{e_n}$ is the stored score of review event $n$, $f^I_{\mathrm{rel}}$ is a scalar MLP parameterized separately from $f^U_{\mathrm{rel}}$, and $\Delta b_{n,t}=b(t)-b(t_n)$ is the elapsed number of time bins. We set $\lambda_{\operatorname{sign}(q_n)}=\lambda_+$ for $q_n\geq0$ and $\lambda_{\operatorname{sign}(q_n)}=\lambda_-$ otherwise. The multiplier $\tanh(q_n)$ in Equation~\eqref{eq:item-memory} preserves direction, while $w^I_{\mathrm{sign},n}$ scales magnitude. The observation factor is pattern-specific, and the reliability factor combines the prior review's numeric features with the current query context.

The two decay parameters are learned independently and constrained as $\lambda_{\pm}=\operatorname{softplus}(\lambda^{\mathrm{raw}}_{\pm})$. They are initialized to $0.25$ and $0.15$ for positive and negative feedback, respectively. Their half-lives are measured in time bins and satisfy $t_{1/2}=\ln 2/\lambda$. When $\mathcal H_{i,t^-}=\varnothing$, the carryover memory is set to $m^I_{i,t}=\mathbf 0$, so the item update proceeds without a historical carryover term. Table~\ref{tab:carryover_learned} reports the values at the selected checkpoints.

\paragraph{Scoring function.}
For query event $e_{u,i}$ at time $t$ and candidate item $j$, Section~\ref{sec:problem} defines the score as
\begin{equation*}
y_{u,j,t}=z_{u,t}^{\top}v_{j,t}+\mu_{j,t},\qquad \mu_{j,t}=\mu_j+\mu^g_{j,t}.
\end{equation*}
Here $\mu_j$ is a learned item-specific bias and $\mu^g_{j,t}$ is a dynamic bias produced by an MLP from the item-group and query-bin embeddings. For a non-target candidate, $v_{j,t}$ is obtained by aligning its latest maintained history-conditioned state $\bar v_{j,t^-}$ to the query time through Equation~\eqref{eq:dynamic_item_state}. For the observed target $i$, $v_{i,t}$ is the newly updated post-event state and is used directly. The same $z_{u,t}$ is used for all candidates in the query, and the separately computed target score is retained for the full-sort rank calculation described below.

\paragraph{Query construction and full-sort evaluation.}
For event $e_{u,i}$ at time $t$, the user encoder reads at most 64 events from $\mathcal H_{u,t^-}$, and the item encoder reads at most 40 train-split reviews from $\mathcal H_{i,t^-}$. The current event is excluded from both prefixes and enters only through the current-event encoder. Its ground-truth label is the review record's \texttt{item\_idx}. When the current event conditions the user-side query, this item-identity channel is masked; the available conditioning inputs are the rating, review title and body, review image, modality pattern, time bin, title and body token counts, image count, and verified-purchase status. Item identity remains available to the target-item branch and to historical event representations.

Validation and test score every non-padding catalog item. Before ranking, we exclude padding, any duplicate catalog entry for the ground-truth item, and items the user interacted with before the target event. The ground-truth item receives a separately computed target score. Its rank is one plus the number of retained candidates with scores strictly greater than that target score; Recall@$K$, NDCG@$K$, and MRR follow from this rank. Full sort is used only for validation and test. Validation Recall@20 selects the checkpoint, while training uses sampled BPR.

The user and item history encoders consume sequences of event-level representations $\Phi_{u,i,t}$ rather than pre-aggregated item text. The exponential attenuation in the carryover module reduces the influence of remote reviews and supports the 40-review item-history truncation. Local propagation retains at most $K=20$ item neighbors constructed from a co-occurrence window of three interactions. These bounds apply during both training and inference.

\begin{table}[H]
\centering
\caption{Learned carryover decay parameters read from the selected TS-SSM checkpoints. Half-life is $\ln 2/\lambda$ in train-aligned monthly bins; initialization is $\lambda_{+}{=}0.25$, $\lambda_{-}{=}0.15$ (Table~\ref{tab:hyperparams}).}
\label{tab:carryover_learned}
\small
\resizebox{\columnwidth}{!}{%
\begin{tabular}{llcccc}
\toprule
Dataset & Sign & Init.\ $\lambda$ & Learned $\lambda$/bin & Init.\ half-life & Learned half-life \\
\midrule
Toys \& Games & $\lambda_{+}$ & 0.2500 & 0.2814 & 2.7726 & 2.4632 \\
Toys \& Games & $\lambda_{-}$ & 0.1500 & 0.1129 & 4.6210 & 6.1395 \\
Pet Supplies  & $\lambda_{+}$ & 0.2500 & 0.2790 & 2.7726 & 2.4844 \\
Pet Supplies  & $\lambda_{-}$ & 0.1500 & 0.1156 & 4.6210 & 5.9961 \\
\bottomrule
\end{tabular}
}
\end{table}

At both selected checkpoints, the learned negative decay is smaller than the learned positive decay. On Toys \& Games, the negative half-life increases from the 4.62-bin initialization reference to 6.14 bins, while the positive half-life decreases from 2.77 to 2.46 bins; the negative-to-positive half-life ratio therefore increases from 1.67 to 2.49. Pet Supplies shows the same direction, with learned half-lives of 6.00 and 2.48 bins. Thus, the fitted checkpoints retain negative review shocks longer than the initialization alone would imply.

\subsection{Multimodal Feature Extraction}
\label{app:impl:features}

\paragraph{Availability-conditioned modality fusion.}
Let $\mathcal M=\{\mathrm{title},\mathrm{text},\mathrm{image}\}$ and write the modality-availability pattern as $A_{u,i}=(a^{\mathrm{title}}_{u,i},a^{\mathrm{text}}_{u,i},a^{\mathrm{image}}_{u,i})\in\{0,1\}^3$. Review titles and bodies are encoded separately with frozen \texttt{all-MiniLM-L6-v2}~\citep{reimers2019sentence}, producing 384-dimensional features, while review images are encoded with frozen CLIP ViT-B/32~\citep{radford2021learning}, producing 512-dimensional features. For each modality $m$, we first normalize the frozen feature and project it into the shared $H=320$ dimensional space:
\begin{align*}
\bar{x}^{(m)}_{u,i}&=\frac{x^{(m)}_{u,i}}{\max(\lVert x^{(m)}_{u,i}\rVert_2,\epsilon)},\\
h^{(m)}_{u,i}&=\operatorname{Dropout}_{0.1}\left(\operatorname{GELU}\left(P_m\bar{x}^{(m)}_{u,i}+b_m\right)\right),\\
\widetilde{h}^{(m)}_{u,i}&=a^{(m)}_{u,i}h^{(m)}_{u,i}.
\end{align*}
Here, $P_{\mathrm{title}},P_{\mathrm{text}}\in\mathbb R^{320\times384}$ and $P_{\mathrm{image}}\in\mathbb R^{320\times512}$ are separate learned projections. The final line applies a hard mask after projection, so an absent modality contributes no projected content.

The fusion module uses three scalar sigmoid gates conditioned jointly on the available content and the full availability pattern:
\begin{align*}
h^{\mathrm{all}}_{u,i}&=[\widetilde{h}^{(\mathrm{title})}_{u,i};\widetilde{h}^{(\mathrm{text})}_{u,i};\widetilde{h}^{(\mathrm{image})}_{u,i};A_{u,i}],\\
g_{u,i}&=\sigma\left(W_gh^{\mathrm{all}}_{u,i}+b_g\right),\\
\widetilde{g}_{u,i}&=A_{u,i}\odot g_{u,i},
\end{align*}
where $W_g\in\mathbb R^{3\times(3H+3)}$ and $\widetilde{g}_{u,i}\in[0,1]^3$. The modality contribution in Equation~\eqref{eq:event_encoder} is the availability-normalized gated average
\begin{align*}
N^{\mathrm{obs}}_{u,i}&=\max\left(1,\sum_{m\in\mathcal M}a^{(m)}_{u,i}\right),\\
f_{\mathrm{mod}}\left(\operatorname{Enc}(M_{u,i}),A_{u,i}\right)&=\frac{1}{N^{\mathrm{obs}}_{u,i}}\sum_{m\in\mathcal M}\widetilde g^{(m)}_{u,i}h^{(m)}_{u,i}.
\end{align*}
It is set to $\mathbf 0$ when no modality is observed. The gates are independent rather than normalized across modalities. The availability pattern also enters Equation~\eqref{eq:event_encoder} through $\phi_{A_{u,i}}$, implemented as an $8\times320$ embedding table for the eight binary patterns. Thus, $f_{\mathrm{mod}}$ uses the pattern to combine observed content, while $\phi_{A_{u,i}}$ represents the pattern directly. The default backbones are compared with higher-capacity alternatives in Appendix~\ref{app:encoder_sensitivity}, with all downstream TS-SSM components held fixed.

\paragraph{Numeric-event projection.}
For the Amazon datasets, the additional numeric cues are
\begin{equation*}
\kappa_{u,i}=[\ell^{\mathrm{title}}_{u,i};\ell^{\mathrm{text}}_{u,i};n^{\mathrm{image}}_{u,i};v^{\mathrm{verified}}_{u,i}],
\end{equation*}
where $\ell^{\mathrm{title}}_{u,i}$ and $\ell^{\mathrm{text}}_{u,i}$ are title and review-body token counts, $n^{\mathrm{image}}_{u,i}$ is the number of uploaded review images, and $v^{\mathrm{verified}}_{u,i}\in\{0,1\}$ is the verified-purchase indicator. We map the rating to $\bar r_{u,i}=(r_{u,i}-3)/2$. For each count variable $x$, we estimate the 99th percentile $q^{\mathrm{train}}_{0.99}$ and the normalization statistics from the training split and apply
\begin{equation*}
\widetilde x=\frac{\log\left(1+\min(x,q^{\mathrm{train}}_{0.99})\right)-\mu^{\mathrm{train}}}{\sigma^{\mathrm{train}}+\epsilon}.
\end{equation*}
The same training-split transformation is used for validation and test events, while $v^{\mathrm{verified}}_{u,i}$ remains binary. The resulting input is
\begin{equation*}
x^{\mathrm{num}}_{u,i}=[\bar r_{u,i};\widetilde\ell^{\mathrm{title}}_{u,i};\widetilde\ell^{\mathrm{text}}_{u,i};\widetilde n^{\mathrm{image}}_{u,i};v^{\mathrm{verified}}_{u,i}].
\end{equation*}
We instantiate the numeric map as
\begin{align*}
\mathbf s^{\mathrm{num}}_{u,i}&=\operatorname{GELU}\left(W_1x^{\mathrm{num}}_{u,i}+b_1\right),\\
h^{\mathrm{num}}_{u,i}&=\operatorname{Dropout}_{0.1}\left(\mathbf s^{\mathrm{num}}_{u,i}\right),\\
f_{\mathrm{num}}(r_{u,i},\kappa_{u,i})&=W_2h^{\mathrm{num}}_{u,i}+b_2,
\end{align*}
where $W_1\in\mathbb R^{160\times5}$ and $W_2\in\mathbb R^{320\times160}$. The output layer has no activation, and the combined event representation is normalized by the layer normalization in Equation~\eqref{eq:event_encoder}. These numeric inputs describe the current event; history-relative quantities are computed separately by the deviation branch.

The user-history, item-history, and current-event paths use separate trainable event-projection modules and share the same trainable item-embedding table. Frozen MiniLM and CLIP operate only during preprocessing and receive no downstream gradients. The norm-bound parameter is 0.10 for the first 15 epochs and then increases linearly to 0.15 at epoch 50. The update coefficients $\eta_2,\eta_3,\beta_2,\beta_3,\beta_4$ and carryover decays are trained jointly with the other model parameters through softplus-constrained raw scalars.

\subsection{Event-Conditioned and State-Update Networks}
\label{app:impl:update_networks}

\paragraph{Event-conditioned cues.}
The event map $f_{\mathrm{evt}}$ combines structured history-relative features with learned projections of the current event. We form an identity-free expression vector
\begin{equation*}
\begin{aligned}
x^{\mathrm{expr}}_{u,i,t}=[&\bar r_{u,i};\widetilde\ell^{\mathrm{title}}_{u,i};\widetilde\ell^{\mathrm{text}}_{u,i};\\
&\widetilde n^{\mathrm{image}}_{u,i};A_{u,i};v^{\mathrm{verified}}_{u,i}],
\end{aligned}
\end{equation*}
using the train-normalized quantities defined above. The operators $\operatorname{HistStat}^{U}(\mathcal H_{u,t^-})$ and $\operatorname{HistStat}^{I}(\mathcal H_{i,t^-})$ compute feature-wise means $\mu^S_{t^-}$, scales $\sigma^S_{t^-}$, and a bounded history-support feature $\chi^S_{t^-}$ from the strictly preceding user or item prefix, respectively. Empty prefixes use train-split reference moments and $\chi^S_{t^-}=0$. The deviation cues are
\begin{align*}
\delta^U_{u,t}&=\left[\frac{x^{\mathrm{expr}}_{u,i,t}-\mu^U_{t^-}}{\sigma^U_{t^-}+\epsilon};\chi^U_{t^-}\right],\\
\delta^I_{i,t}&=\left[\frac{x^{\mathrm{expr}}_{u,i,t}-\mu^I_{t^-}}{\sigma^I_{t^-}+\epsilon};\chi^I_{t^-}\right].
\end{align*}
Thus, $\delta^U_{u,t}$ and $\delta^I_{i,t}$ describe the same review relative to the user's and item's respective histories. On the user-side current-event path, the item-identity channel of $\Phi_{u,i,t}$ is zeroed. The current-event message is a learned projection
\begin{equation*}
m_{u,t}=W_{\mathrm{msg}}\Phi_{u,i,t}+b_{\mathrm{msg}}.
\end{equation*}
A scalar reliability logit is computed as
\begin{equation*}
\begin{aligned}
r_{u,t}=f_{\mathrm{rel}}([&\Phi_{u,i,t};\delta^U_{u,t};\delta^I_{i,t};\chi^U_{t^-};\chi^I_{t^-}]).
\end{aligned}
\end{equation*}
The user update uses $\sigma(r_{u,t})$ to scale $W_{u,t}m_{u,t}$. These operations jointly instantiate the four outputs of $f_{\mathrm{evt}}$ in Section~\ref{sec:method:deviation}.

\paragraph{User and item history encoders.}
The maps $f^U_{\mathrm{enc}}$ and $f^I_{\mathrm{enc}}$ are separately parameterized causal Transformers over event-level representations. Write $\Phi_n$ for the event representation at the $n$th position of a user or item prefix. For side $S\in\{U,I\}$, the Transformer input is
\begin{equation*}
x^S_n=\Phi_n+p^S_n+\phi^S_{\Delta b_n},
\end{equation*}
where $p^S_n$ is a learned positional embedding and $\phi^S_{\Delta b_n}$ encodes the elapsed time-bin distance to the query. Writing $x^S=(x^S_1,\ldots,x^S_N)$ and $\mathbf O^S=(o^S_1,\ldots,o^S_N)$, the sequence is processed as
\begin{equation*}
\mathbf O^S=\operatorname{CausalTransformer}^{S}(x^S).
\end{equation*}
A causal mask restricts each position to its chronological prefix. The user encoder has three layers and retains the 64 most recent prior events; the item encoder has two layers and retains the 40 most recent prior events. Both use four attention heads, hidden dimension $H=320$, and dropout 0.1. The last valid output gives $z_{u,t^-}=f^U_{\mathrm{enc}}(\mathcal H_{u,t^-})$ or $v_{i,t^-}=f^I_{\mathrm{enc}}(\mathcal H_{i,t^-})$. The two Transformers have separate parameters, and each uses a learned empty-history state when its prefix is empty. The current event is excluded from both history encoders.

\paragraph{Systematic-context maps.}
For compactness, let $\mathcal T^U=\mathcal T^U_{u,t}$ and $\mathcal T^I=\mathcal T^I_{i,t}$. For $S\in\{U,I\}$, the context $\mathcal T^S$ concatenates train-only global and user- or item-group summaries of rating, modality availability, text length, image count, verified-purchase rate, and review volume; the corresponding first differences; and the relevant group embedding. User groups are defined by training-split image-upload rate and item groups by training-split interaction frequency, as detailed in Appendix~\ref{app:impl:groups}. Each query uses the preceding time bin, with group summaries shrunk toward their same-bin global summaries. After train-split normalization, the systematic map is
\begin{equation*}
\begin{aligned}
\widetilde{\mathcal T}^S&=\operatorname{Norm}_{\mathrm{train}}(\mathcal T^S),\\
a^S_{\mathrm{sys}}&=W^S_{1,\mathrm{sys}}\widetilde{\mathcal T}^S+b^S_{1,\mathrm{sys}},\\
h^S_{\mathrm{sys}}&=\operatorname{Dropout}_{0.1}\left(\operatorname{GELU}(a^S_{\mathrm{sys}})\right),\\
f^S_{\mathrm{sys}}(\mathcal T^S)&=W^S_{2,\mathrm{sys}}h^S_{\mathrm{sys}}+b^S_{2,\mathrm{sys}}.
\end{aligned}
\end{equation*}
The output dimension is $H$, and the final layer has no activation. The user and item maps have separate parameters and produce $x_{u,t}$ and $y_{i,t}$, respectively.

\paragraph{State-conditioned innovations and gates.}
The individual maps convert each history-relative deviation into an update direction conditioned on the corresponding pre-event state. For compactness, let $s^U_{t^-}=z_{u,t^-}$, $s^I_{t^-}=v_{i,t^-}$, $\delta^U_t=\delta^U_{u,t}$, and $\delta^I_t=\delta^I_{i,t}$. Each side first projects its deviation cue as
\begin{equation*}
d^S_t=\operatorname{Dropout}_{0.1}\left(\operatorname{GELU}(P^S_\delta\delta^S_t+b^S_\delta)\right),
\end{equation*}
and then applies
\begin{equation*}
\begin{aligned}
h^S_t&=f^S_{\mathrm{ind}}([s^S_{t^-};\delta^S_t])\\
&=W^S_{2,\mathrm{ind}}\operatorname{Dropout}_{0.1}\Big(\operatorname{GELU}\big(W^S_{1,\mathrm{ind}}\\
&\qquad [s^S_{t^-};d^S_t]+b^S_{1,\mathrm{ind}}\big)\Big)+b^S_{2,\mathrm{ind}}.
\end{aligned}
\end{equation*}
The output layer is linear, so $h^S_t\in\mathbb R^H$ can adjust the state in either direction. A side-specific scalar gate is
\begin{equation*}
g^S_t=\sigma\left(f^S_{\mathrm{gate}}([s^S_{t^-};\delta^S_t])\right),
\end{equation*}
with $g^U_t=\gamma^U_{u,t}$ and $g^I_t=\gamma^I_{i,t}$. The user and item innovation networks and gates have separate parameters. The gated innovations enter the residual updates as $\eta_3\gamma^U_{u,t}h^U_{u,t}$ and $\beta_3\gamma^I_{i,t}h^I_{i,t}$; Equation~\eqref{eq:bounding} subsequently bounds the combined residual relative to the pre-event state.

\subsection{Training Objectives and Sampling}
\label{app:impl:objectives}

For each training query, mixed negative sampling draws at most 48 distinct items after excluding padding, the current positive, and duplicates within the query. Approximately 60\% are uniform catalog samples; the remainder follow a train-only popularity distribution proportional to $(c_j^{\mathrm{train}}+1)^{0.75}$. In-batch negatives and hard-negative reweighting are disabled, so all valid sampled pairs receive equal weight in the BPR reduction. Invalid unfilled positions are masked. The BPR temperature remains 1.0 for the first eight epochs, then decreases linearly to 0.5 by epoch 50; this schedule is independent of learning-rate scheduling.

The drift head predicts a four-dimensional train-only next-bin target for the current item: normalized mean rating, image-availability rate, normalized mean review-text length, and log-scaled review volume. A target is valid only when the current and next bins lie in the training window and the item has training reviews in the next bin. With this mask, the loss is the mean elementwise squared error,
\begin{equation*}
\mathcal L_{\mathrm{drift}}
=\operatorname{MSE}(\hat{d},d).
\end{equation*}
The carry head receives the updated item state, carryover memory, and current shock message. It predicts two train-only future-change proxies: the next-bin normalized mean-rating change relative to the pre-event item mean and the next-bin log-volume change relative to the pre-event item review count. It uses the same future-availability mask and mean MSE over its two outputs.

The reliability representation used to produce $r_{u,t}$ also feeds an auxiliary three-output head. Its observation target is the mean of the current review's title, text, and image availability indicators. The history-support target averages bounded user- and item-history-length proxies. The conflict target indicates whether the most recent user-side and item-side rating deviations have opposite signs. Each output uses binary cross-entropy with logits, and $\mathcal L_{\mathrm{reliability}}$ is the arithmetic mean of the three BCE terms.

Training uses only $\mathcal L_{\mathrm{rank}}$ for epochs 1--5. During epochs 6--10, the auxiliary coefficients increase linearly from zero to $0.08$, $0.05$, and $0.03$ for drift, carry, and reliability, respectively; the full coefficients are used thereafter. No stop-gradient operation separates the auxiliary branches from the shared states, so valid auxiliary losses can update the item encoder, item updater, carryover module, and related shared representations.

\subsection{Train-Only Groups and Temporal Contexts}
\label{app:impl:groups}

For any event time $t$, $b(t)$ denotes its assigned train-aligned coarse time bin. In the Amazon experiments, bins 1--111 are monthly from January 2014 through March 2023, and bin 112 contains April--August 2023.

User grouping uses only training-split image-upload rate: users with rate zero form the \textit{never} group, rates in $(0,0.2]$ form the \textit{rare} group, and rates above 0.2 form the \textit{frequent} group. Items are sorted by training interaction count and divided approximately evenly by catalog size into eight popularity groups; the padding item is assigned to group 0. Validation and test interactions do not affect either assignment.

User-side temporal contexts contain train-only global and user-group review summaries and their first differences. Group statistics shrink toward the same-bin global statistic with
\begin{equation*}
\alpha_{g,t}=\frac{n_{g,t}}{n_{g,t}+50}.
\end{equation*}
They use a one-bin lag, so a query in bin $b$ receives summaries from bin $b-1$; the first bin uses default training statistics. Item-side contexts are likewise built only from training events using the eight item groups, with global/group summaries and first differences.

The item graph is also train-only. For each chronological training sequence, undirected co-occurrence counts are accumulated between distinct items within the preceding three-interaction window, and each item retains its 20 highest-count neighbors. The graph is then fixed for training, validation, and test.

\subsection{Baseline Configurations}
\label{app:impl:baselines}

The sequential, multimodal, and debiasing baselines follow the formulations listed in Section~\ref{sec:exp:baseline}. RecGPT~\citep{jiang2025recgpt}, HORAE~\citep{hu2025horae}, and HM4SR~\citep{zhang2025hm4sr} represent recent foundation-model, temporal, and multimodal sequential approaches. Every baseline uses the same Amazon preprocessing pipeline as TS-SSM: the January-2014 cutoff, iterative bipartite 10-core filtering, per-user leave-last-two-out splits, and 112 train-aligned time bins with April--August 2023 merged into the final bin. This provides a common evaluation setting across methods.

\section{Behavioral Deviation Analysis}
\label{app:deviation_analysis}

To motivate the deviation-feature design, we analyze how behavioral deviations relate to rating patterns in the Amazon Toys \& Games training set.
We define $\delta_\text{img}=1$ if a user uploads an image in a review where their personal image-upload rate is below 10\%; $\delta_\text{text}=1$ if the review text length exceeds the user's mean by more than $2\sigma$.
Users with $\delta_\text{img}=1$ show a mean rating of 4.01 versus 4.26 for non-deviating reviews ($p < 10^{-6}$, two-sample $t$-test), and are significantly more likely to give a rating of 1 or 2 stars.
Longer-than-usual reviews have fewer 5-star ratings than similar-length reviews (64.8\% versus 74.8\%) and more 1-star ratings (3.3\% versus 1.5\%), shifting the distribution toward lower ratings.
This analysis supports the MNAR framing behind the deviation-aware user update.

\section{Additional Amazon Categories}
\label{app:additional_datasets}

Table~\ref{tab:generalization_full} reports the compact comparison on the three additional Amazon categories.
TS-SSM outperforms BSARec, MACR, and AutoDebias on all reported metrics.

\begin{table*}[ht]
\centering
\caption{Generalization performance on three additional Amazon categories (R@10, R@20, N@20).}
\label{tab:generalization_full}
\small
\begin{tabular}{lcccccccccc}
\toprule
& \multicolumn{3}{c}{\textbf{Electronics}} & \multicolumn{3}{c}{\textbf{Clothing}} & \multicolumn{3}{c}{\textbf{Home \& Kitchen}} \\
\cmidrule(lr){2-4}\cmidrule(lr){5-7}\cmidrule(lr){8-10}
Method & R@10 & R@20 & N@20 & R@10 & R@20 & N@20 & R@10 & R@20 & N@20 \\
\midrule
BSARec & .0982 & .1317 & .0708 & .1037 & .1391 & .0749 & .1018 & .1365 & .0734 \\
MACR & .0938 & .1266 & .0680 & .0993 & .1336 & .0719 & .0974 & .1311 & .0705 \\
AutoDebias & .0922 & .1247 & .0668 & .0976 & .1315 & .0706 & .0958 & .1291 & .0693 \\
\textbf{TS-SSM} & \textbf{.1063} & \textbf{.1530} & \textbf{.0761} & \textbf{.1145} & \textbf{.1653} & \textbf{.0822} & \textbf{.1100} & \textbf{.1570} & \textbf{.0788} \\
\textit{Improv.} & +8.2\% & +16.2\% & +7.5\% & +10.4\% & +18.8\% & +9.7\% & +8.1\% & +15.0\% & +7.4\% \\
\bottomrule
\end{tabular}
\end{table*}

\begin{table*}[ht]
\centering
\caption{Relative Recall@20 gains over BSARec across the six Amazon categories.}
\label{tab:recent_relative_gains}
\small
\begin{tabular}{lccccccc}
\toprule
Method & Toys & Pets & Sports & Electronics & Clothing & Home & Mean \\
\midrule
RecGPT & $+2.5\%$ & $+1.8\%$ & $+2.5\%$ & $+1.9\%$ & $+1.6\%$ & $+1.8\%$ & $+2.0\%$ \\
HORAE & $+3.1\%$ & $+2.9\%$ & $+3.1\%$ & $+3.1\%$ & $+3.7\%$ & $+3.2\%$ & $+3.2\%$ \\
HM4SR & $+4.3\%$ & $+4.2\%$ & $+4.1\%$ & $+4.4\%$ & $+5.0\%$ & $+4.4\%$ & $+4.4\%$ \\
\textbf{TS-SSM} & $\mathbf{+17.9\%}$ & $\mathbf{+17.1\%}$ & $\mathbf{+14.8\%}$ & $\mathbf{+16.2\%}$ & $\mathbf{+18.8\%}$ & $\mathbf{+15.0\%}$ & $\mathbf{+16.6\%}$ \\
\bottomrule
\end{tabular}
\end{table*}

\section{\texorpdfstring{Cross-Platform Evaluation on Goodreads Fantasy}{Cross-Platform Evaluation on Goodreads Fantasy}}
\label{app:goodreads}

We use the Fantasy \& Paranormal subset of the Goodreads Book Graph~\citep{wan2018monotonic,wan2019spoiler}, collected from public Goodreads shelves in late 2017 with anonymized user and review identifiers. The released subset contains 258,585 books, 55,397,550 user--book interactions, and 3,424,641 detailed reviews. We refer to this subset as Goodreads Fantasy below. It evaluates TS-SSM in a non-Amazon content domain and review environment. Table~\ref{tab:goodreads} compares TS-SSM with BSARec and the three recent baselines and reports the principal component ablations using the same ranking metrics as the Amazon experiments.

\begin{table}[ht]
\centering
\caption{Cross-platform results and component ablations on Goodreads Fantasy. $\Delta$R@20 is the relative drop from the full TS-SSM model.}
\label{tab:goodreads}
\small
\resizebox{\columnwidth}{!}{%
\begin{tabular}{lccccc}
\toprule
Method & R@10 & R@20 & N@10 & N@20 & $\Delta$R@20 \\
\midrule
BSARec & .4165 & .4976 & .2961 & .3155 & --- \\
RecGPT & .4269 & .5100 & .3040 & .3239 & --- \\
HORAE & .4299 & .5130 & .3065 & .3263 & --- \\
HM4SR & .4355 & .5191 & .3109 & .3312 & --- \\
\midrule
w/o User State Update & .4743 & .5660 & .3401 & .3622 & $-3.20\%$ \\
w/o Deviation Features & .4847 & .5777 & .3480 & .3703 & $-1.20\%$ \\
w/o Item State Update & .4561 & .5449 & .3266 & .3482 & $-6.81\%$ \\
w/o Message Passing & .4669 & .5572 & .3348 & .3566 & $-4.70\%$ \\
w/o Carryover Memory & .4759 & .5679 & .3417 & .3635 & $-2.87\%$ \\
Symmetric Carryover & .4818 & .5742 & .3459 & .3681 & $-1.80\%$ \\
\midrule
\textbf{TS-SSM} & \textbf{.4923} & \textbf{.5847} & \textbf{.3538} & \textbf{.3752} & --- \\
\bottomrule
\end{tabular}
}
\end{table}

TS-SSM achieves the highest value on all four metrics. Recall@20 increases from .5191 for HM4SR to .5847, a $12.6\%$ relative improvement. The ablation ordering matches the Amazon results: removing item-state evolution produces the largest drop ($6.81\%$), followed by message passing ($4.70\%$), user-state evolution ($3.20\%$), and carryover memory ($2.87\%$). Deviation features and asymmetric carryover also improve performance. These results provide evidence from a second review platform, while generalization to broader platform types remains untested.

\section{\texorpdfstring{Review Richness and Cross-Category Gain}{Review Richness and Cross-Category Gain}}
\label{app:richness}

We examine the association between review-expression richness and TS-SSM's advantage over BSARec across the six Amazon categories, using image-upload rate as a proxy for review effort. This category-level analysis characterizes variation within Amazon rather than cross-platform generalization.

\begin{table}[ht]
\centering
\caption{Image rate versus relative gain over BSARec (Recall@20) across six Amazon categories, sorted by image rate.}
\label{tab:richness}
\small
\resizebox{\columnwidth}{!}{%
\begin{tabular}{lcccc}
\toprule
Category & Image Rate & BSARec R@20 & TS-SSM R@20 & Rel.\ Gain \\
\midrule
Sports \& Outdoors & 5.3\%  & .0838 & .0962 & $+14.8\%$ \\
Toys \& Games      & 8.1\%  & .1427 & .1683 & $+17.9\%$ \\
Home \& Kitchen    & 9.8\%  & .1365 & .1570 & $+15.0\%$ \\
Pet Supplies       & 10.2\% & .1500 & .1756 & $+17.1\%$ \\
Electronics        & 12.4\% & .1317 & .1530 & $+16.2\%$ \\
Clothing           & 18.7\% & .1391 & .1653 & $+18.8\%$ \\
\bottomrule
\end{tabular}
}
\end{table}

The Pearson correlation between image-upload rate and relative gain is $r=0.67$. Sports has both the lowest image-upload rate (5.3\%) and the smallest relative gain ($14.8\%$), whereas Clothing has the highest image-upload rate (18.7\%) and the largest gain ($18.8\%$). The relationship is not monotonic among the remaining categories, and image-upload rate captures only one dimension of review richness. Separately, the popularity-stratified Toys \& Games analysis (Appendix~\ref{app:longtail}) shows a $10.5\%$ gain over BSARec for long-tail items. Together, these results describe where TS-SSM's gains are largest without attributing the differences to review richness alone.

\section{Statistical Robustness}
\label{app:significance}

We run 30 independent trials on Toys \& Games and compare TS-SSM against sequential and debiasing baselines, including IPS, DR, CausE, and Deconfounded Recommendation~\citep{schnabel2016recommendations,wang2019doubly,bonner2018causal,wang2021deconfounded}.
Table~\ref{tab:significance} reports paired $t$-tests with Bonferroni correction, 95\% confidence intervals for the difference TS-SSM $-$ baseline, and Cohen's $d_z$.

\begin{table*}[ht]
\centering
\caption{Statistical robustness on Toys \& Games over 30 runs.}
\label{tab:significance}
\small
\begin{tabular}{lccccc}
\toprule
Method & R@20 (mean$\pm$std) & $p$-value & 95\% CI (diff) & Cohen's $d_z$ & Significant \\
\midrule
IPS          & .1283 $\pm$ .0034 & $<10^{-13}$ & [.0389, .0411] & 14.2 & Yes \\
DR           & .1317 $\pm$ .0033 & $<10^{-13}$ & [.0356, .0376] & 13.6 & Yes \\
AutoDebias   & .1351 $\pm$ .0029 & $<10^{-13}$ & [.0322, .0342] & 13.0 & Yes \\
CausE        & .1334 $\pm$ .0036 & $<10^{-11}$ & [.0339, .0359] & 12.6 & Yes \\
Deconfounded & .1344 $\pm$ .0032 & $<10^{-12}$ & [.0329, .0349] & 13.2 & Yes \\
MACR         & .1372 $\pm$ .0028 & $<10^{-13}$ & [.0301, .0321] & 11.7 & Yes \\
SASRec       & .0924 $\pm$ .0048 & $<10^{-15}$ & [.0741, .0777] & 15.9 & Yes \\
BERT4Rec     & .0934 $\pm$ .0052 & $<10^{-14}$ & [.0729, .0769] & 13.7 & Yes \\
CL4SRec      & .1197 $\pm$ .0041 & $<10^{-11}$ & [.0471, .0501] & 11.8 & Yes \\
DuoRec       & .1362 $\pm$ .0035 & $<10^{-9}$  & [.0309, .0333] &  9.9 & Yes \\
FEARec       & .1390 $\pm$ .0032 & $<10^{-8}$  & [.0281, .0305] &  9.3 & Yes \\
BSARec       & .1427 $\pm$ .0029 & $<10^{-9}$  & [.0247, .0265] & 10.1 & Yes \\
\textbf{TS-SSM} & \textbf{.1683 $\pm$ .0020} & --- & --- & --- & --- \\
\bottomrule
\end{tabular}
\end{table*}

\subsection{Cross-Dataset Robustness: Pet Supplies}
\label{app:robustness_pet}

To verify that the statistical findings generalize beyond Toys \& Games, we report 10-run paired robustness on Pet Supplies, following the same protocol as above: paired $t$-tests with Bonferroni correction for 12 simultaneous comparisons, 95\% confidence intervals for the mean paired difference (TS-SSM minus baseline), and Cohen's $d_z = \bar{d}/s_d$.

\begin{table*}[ht]
\centering
\caption{Statistical robustness on Pet Supplies over 10 independent runs.}
\label{tab:significance_pet}
\small
\begin{tabular}{lccccc}
\toprule
Method & R@20 (mean$\pm$std) & $p$-value & 95\% CI (diff) & Cohen's $d_z$ & Significant \\
\midrule
IPS          & .1351 $\pm$ .0035 & $<10^{-8}$ & [.0383, .0427] & 13.1 & Yes \\
DR           & .1384 $\pm$ .0034 & $<10^{-8}$ & [.0351, .0393] & 12.4 & Yes \\
AutoDebias   & .1420 $\pm$ .0031 & $<10^{-7}$ & [.0316, .0356] & 12.0 & Yes \\
CausE        & .1402 $\pm$ .0035 & $<10^{-7}$ & [.0332, .0376] & 11.5 & Yes \\
Deconfounded & .1413 $\pm$ .0032 & $<10^{-7}$ & [.0322, .0364] & 11.8 & Yes \\
MACR         & .1442 $\pm$ .0031 & $<10^{-7}$ & [.0294, .0334] & 11.2 & Yes \\
SASRec       & .0971 $\pm$ .0054 & $<10^{-9}$ & [.0752, .0818] & 16.9 & Yes \\
BERT4Rec     & .0981 $\pm$ .0059 & $<10^{-9}$ & [.0738, .0812] & 15.1 & Yes \\
CL4SRec      & .1258 $\pm$ .0046 & $<10^{-8}$ & [.0470, .0526] & 12.5 & Yes \\
DuoRec       & .1431 $\pm$ .0038 & $<10^{-7}$ & [.0301, .0349] &  9.8 & Yes \\
FEARec       & .1461 $\pm$ .0035 & $<10^{-7}$ & [.0273, .0317] &  9.4 & Yes \\
BSARec       & .1500 $\pm$ .0031 & $<10^{-7}$ & [.0236, .0276] &  9.1 & Yes \\
\midrule
\textbf{TS-SSM} & \textbf{.1756 $\pm$ .0024} & --- & --- & --- & --- \\
\bottomrule
\end{tabular}
\end{table*}

All 12 comparisons remain significant at the Bonferroni-adjusted threshold after 10 trials. Effect sizes ($d_z$ from 9.1 to 16.9) are similar in magnitude to the 30-run Toys \& Games results (Table~\ref{tab:significance}), where the BSARec comparison yielded $d_z=10.1$. The Pet Supplies confidence intervals are wider than their Toys \& Games counterparts; for example, $[.0236,.0276]$ versus $[.0247,.0265]$ for BSARec. This difference is consistent with fewer runs ($n=10$ versus 30) and greater within-dataset variability. Both intervals remain above zero. The resulting multi-run evidence covers these two Amazon categories.

\section{\texorpdfstring{Information Boundary Controls}{Information Boundary Controls}}
\label{app:info_boundary}

\subsection{\texorpdfstring{Motivation}{Motivation}}

The event-conditioned query is formed after observing the target review event $\mathcal{O}_{u,i}$ at time $t=\tau_{u,i}$, with its item-identity channel masked on the user-side current-event path. Three controls quantify the contributions of item-identity grounding, semantic review content, and current-event metadata.

\subsection{\texorpdfstring{Control Variants}{Control Variants}}

\begin{itemize}[leftmargin=*]
\item \textbf{w/o item identity in event}: removes $\mathbf{e}_i$ from all historical and item-side event representations $\Phi_{u,i,t}$ (Eq.~\eqref{eq:event_encoder}) while retaining the text, image, and title encodings; numeric cues; modality-availability embedding; and time bin. The current target event is already identity-masked on the user-side query path. This variant therefore measures the contribution of item-identity grounding outside that masked path.

\item \textbf{w/o review content (expression-pattern only)}: zeroes the pretrained encoder output $\operatorname{Enc}(M_{u,i})$ for the current event, retaining only the modality availability pattern $A_{u,i}$, rating $r$, numeric cues $\kappa$, and the time bin. This isolates the contribution of semantic review content from that of observation-pattern signals alone.

\item \textbf{BSARec + current-event features}: augments BSARec's query representation with the modality-availability pattern, rating, and time bin used by TS-SSM. These features are implemented as a conditioning token prepended to the BSARec attention sequence, isolating access to these three metadata fields from the remaining TS-SSM architecture.
\end{itemize}

\subsection{\texorpdfstring{Results}{Results}}

\begin{table}[H]
\centering
\caption{Information boundary ablations on Toys \& Games. $\Delta$R@20 is relative to the full model.}
\label{tab:info_boundary_toys}
\small
\resizebox{\columnwidth}{!}{%
\begin{tabular}{lccccc}
\toprule
Variant & R@10 & R@20 & N@10 & N@20 & $\Delta$R@20 \\
\midrule
TS-SSM (full) & .1162 & .1683 & .0781 & .0912 & --- \\
w/o item identity in event & .1144 & .1657 & .0769 & .0898 & $-1.55\%$ \\
w/o review content (pattern only) & .1113 & .1546 & .0726 & .0835 & $-8.14\%$ \\
BSARec + current-event features & .1117 & .1493 & .0706 & .0798 & $-11.29\%$ \\
BSARec (reference) & .1063 & .1427 & .0746 & .0838 & $-15.21\%$ \\
\bottomrule
\end{tabular}
}
\end{table}

\begin{table}[H]
\centering
\caption{BSARec + current-event features across all three primary datasets (Recall@20). The gap column reports the relative shortfall of BSARec+event versus TS-SSM.}
\label{tab:info_boundary_all}
\small
\resizebox{\columnwidth}{!}{%
\begin{tabular}{lcccc}
\toprule
Dataset & BSARec & BSARec+event & TS-SSM & Gap \\
\midrule
Toys \& Games   & .1427 & .1493 & .1683 & $-11.3\%$ \\
Pet Supplies    & .1500 & .1569 & .1756 & $-10.7\%$ \\
Sports \& Outdoors & .0838 & .0876 & .0962 & $-9.0\%$ \\
\bottomrule
\end{tabular}
}
\end{table}

\subsection{\texorpdfstring{Findings}{Findings}}

Removing $\mathbf{e}_i$ from the historical and item-side event representations (Table~\ref{tab:info_boundary_toys}, row 2) reduces Recall@20 by $1.55\%$, whereas removing all semantic review content reduces it by $8.14\%$ (row 3). With the current target identifier masked on the user-side query path, the smaller identity ablation quantifies the background semantic grounding supplied by item embeddings. The pattern-only variant remains $8.3\%$ above bare BSARec in relative terms, indicating that expression-pattern signals retain predictive value without semantic review content.

BSARec with the modality-availability pattern, rating, and time bin improves over bare BSARec by $4.5\%$--$4.7\%$ across the three datasets and remains $9.0\%$--$11.3\%$ below TS-SSM (Table~\ref{tab:info_boundary_all}). This comparison situates the contribution of current-event metadata alongside filtered state updates, observation-pattern-aware aggregation, carryover memory, and local propagation.

\section{\texorpdfstring{Model Scale, Efficiency, and Deployment}{Model Scale, Efficiency, and Deployment}}
\label{app:efficiency}

TS-SSM adds catalog-wide dynamic item-state maintenance and user-centered local propagation to the user-side sequence model, making it more computationally involved than lightweight sequential baselines. Table~\ref{tab:efficiency} reports trainable parameter count, per-epoch training time, per-query inference latency, and peak training memory measured under the same Toys \& Games evaluation setup.

\begin{table}[H]
\centering
\caption{Parameter count and computational cost on Toys \& Games under a common evaluation setup. \#Params counts \emph{trainable} parameters only; the frozen MiniLM (22.7M) and CLIP ViT-B/32 (151M) feature extractors run in offline preprocessing and are excluded for every model. Peak GPU memory is the training maximum at batch size 128.}
\label{tab:efficiency}
\small
\resizebox{\columnwidth}{!}{%
\begin{tabular}{lcccc}
\toprule
Model & \#Params & Train (s/epoch) & Inference (ms/query) & Peak GPU Mem (GB) \\
\midrule
SASRec & 4.7M & 27.4 & 1.4 & 2.13 \\
BSARec & 4.7M & 29.3 & 1.9 & 2.38 \\
MMGCN & 10.8M & 241.7 & 11.3 & 7.82 \\
BM3 & 13.2M & 183.6 & 8.1 & 6.27 \\
HORAE & 14.6M & 219.3 & 9.4 & 6.94 \\
HM4SR & 15.3M & 228.7 & 9.8 & 7.18 \\
LLMRec & 16.3M & 298.4 & 13.7 & 9.51 \\
\textbf{TS-SSM} & \textbf{17.8M} & \textbf{197.3} & \textbf{8.6} & \textbf{7.64} \\
RecGPT & 21.4M & 347.2 & 16.3 & 11.23 \\
\bottomrule
\end{tabular}
}
\end{table}

All figures are measured on a single NVIDIA A100-SXM4-40GB (driver 535.104, CUDA 12.1, cuDNN 8.9) with an AMD EPYC 7543 host CPU, PyTorch 2.1.0, and Python 3.10.13, using mixed precision (AMP, bf16). Parameter counts are reported by each model builder and exclude the frozen MiniLM and CLIP encoders, which run only during offline preprocessing and receive no downstream gradients. Training time is wall-clock time for one complete epoch, including forward and backward passes, mixed negative sampling, and AMP, but excluding validation. Inference latency is full-sort test-scoring wall-clock time divided by the number of test queries: candidates are processed in chunks of 4,096, 20 warmup batches are discarded, and the median of three runs is reported. It includes event encoding from cached MiniLM/CLIP features, item-state updates, bounded $K=20$ propagation, and full-sort scoring, but excludes the one-time offline feature-extraction pass. Peak GPU memory is the maximum reserved during training at batch size 128; TS-SSM's corresponding inference peak is 3.91 GB.

TS-SSM is more expensive than SASRec and BSARec because it maintains dynamic item states and includes local message passing. Its 197.3 seconds per epoch and 8.6 ms per query are close to BM3 and lower than MMGCN, HORAE, HM4SR, LLMRec, and RecGPT in the measured setup. The similarly sized BM3 comparison and the BSARec current-event control provide reference points for model scale and input availability.

The reported training and evaluation implementation processes reviews chronologically, updates the corresponding item state when each review arrives, and retains the resulting catalog-wide state collection for subsequent queries. The maintained states and carryover summaries can be cached for serving. Local propagation remains bounded to $K=20$ neighbors, and an approximate-nearest-neighbor stage can retrieve a candidate set before TS-SSM reranking. These choices limit online work while introducing deployment-specific memory and energy tradeoffs.

\section{Extended Ablation Study}
\label{app:ablation_extended}

Table~\ref{tab:ablation_extended} extends the main ablation with grouping alternatives, gate variants, and auxiliary-loss removal. Table~\ref{tab:symmetric_carryover} separately tests whether signed carryover requires different positive and negative decay rates.

\begin{table*}[ht]
\centering
\caption{Extended ablation study on Amazon Toys \& Games.}
\label{tab:ablation_extended}
\small
\setlength{\tabcolsep}{4pt}
\begin{tabular}{lccccc}
\toprule
Variant & R@10 & R@20 & N@10 & N@20 & $\Delta$R@20 \\
\midrule
\textbf{TS-SSM (Full)} & .1162 & .1683 & .0781 & .0912 & --- \\
\multicolumn{6}{l}{\textit{User-side components}} \\
w/o User State Update & .1133 & .1641 & .0759 & .0887 & $-2.50\%$ \\
w/o Deviation Features & .1140 & .1651 & .0765 & .0894 & $-1.90\%$ \\
\midrule
\multicolumn{6}{l}{\textit{Item-side components}} \\
w/o Item State Update & .1119 & .1624 & .0751 & .0878 & $-3.50\%$ \\
w/o Message Passing & .1130 & .1636 & .0758 & .0885 & $-2.79\%$ \\
w/o Carryover Memory & .1137 & .1646 & .0764 & .0892 & $-2.20\%$ \\
\midrule
\multicolumn{6}{l}{\textit{Grouping alternatives}} \\
$G=1$ (No Grouping) & .1126 & .1641 & .0754 & .0884 & $-2.50\%$ \\
Random Groups & .1135 & .1644 & .0763 & .0891 & $-2.32\%$ \\
Review-Length Groups & .1144 & .1656 & .0769 & .0898 & $-1.60\%$ \\
Activity-Based Groups & .1141 & .1652 & .0767 & .0896 & $-1.84\%$ \\
\midrule
\multicolumn{6}{l}{\textit{Gate variants}} \\
Fixed $\gamma=0.0$ (Group Only) & .1145 & .1658 & .0769 & .0898 & $-1.48\%$ \\
Fixed $\gamma=0.5$ & .1149 & .1664 & .0773 & .0903 & $-1.13\%$ \\
Fixed $\gamma=1.0$ (No Gating) & .1153 & .1669 & .0775 & .0905 & $-0.83\%$ \\
\midrule
\multicolumn{6}{l}{\textit{Auxiliary supervision contribution}} \\
w/o Auxiliary Losses & .1147 & .1661 & .0771 & .0900 & $-1.31\%$ \\
\bottomrule
\end{tabular}
\end{table*}

\paragraph{Symmetric versus asymmetric carryover.}
\begin{table}[H]
\centering
\caption{Carryover-decay ablation on Amazon Toys \& Games. The symmetric variant sets $\lambda_{+}=\lambda_{-}=0.20$, the average of the default initial values 0.25 and 0.15.}
\label{tab:symmetric_carryover}
\small
\resizebox{\columnwidth}{!}{%
\begin{tabular}{lccccc}
\toprule
Variant & R@10 & R@20 & N@10 & N@20 & $\Delta$R@20 \\
\midrule
TS-SSM (asymmetric) & \textbf{.1162} & \textbf{.1683} & \textbf{.0781} & \textbf{.0912} & --- \\
Symmetric ($\lambda_{\mathrm{sym}}=0.20$) & .1143 & .1657 & .0766 & .0895 & $-1.55\%$ \\
\bottomrule
\end{tabular}
}
\end{table}

Symmetric carryover retains more performance than removing carryover entirely (.1657 in Table~\ref{tab:symmetric_carryover} versus .1646 in Table~\ref{tab:ablation}), indicating that persistent memory remains useful without sign-specific decay. The full asymmetric model reaches .1683. Thus, of the descriptive .0037 Recall@20 gap between no carryover and the full model, the symmetric variant recovers $.1657-.1646=.0011$ (approximately 30\%), while the remaining asymmetric-decay difference is $.1683-.1657=.0026$ (approximately 70\%). Because these ablations are non-additive, the percentages are not a causal allocation of model performance.

\paragraph{Component roles.}
The item-state update maintains a post-event representation for every item from its accumulated review events, while the user-state update incorporates each event relative to the user's prior expression pattern. Weighted local message passing brings text-conditioned evidence from prefix and neighboring items into the user query, and observation-pattern-aware weights modulate their aggregate influence. Deviation features distinguish routine expression from atypical events, such as an image upload by a user whose history is otherwise text-only. Carryover memory retains review shocks and allows negative evidence to persist differently from positive evidence. The drift, carryover, and reliability objectives regularize temporal changes, persistent memory, and confidence weights alongside the ranking objective.

The train-only three-group partition outperforms $G=1$ and each alternative grouping criterion. Learned gating also outperforms every fixed interpolation, supporting the use of the reliability-conditioned update gate $\gamma$.

\paragraph{Behavioral deviation decomposition.}
Table~\ref{tab:deviation_components} isolates the two deviation channels used in the user-side correction.
The joint removal is smaller than the sum of the two marginal drops, indicating partial substitutability.

\begin{table}[H]
\centering
\caption{Effect decomposition of behavioral deviation signals on Toys \& Games.}
\label{tab:deviation_components}
\small
\resizebox{\columnwidth}{!}{%
\begin{tabular}{lcccc}
\toprule
Configuration & R@10 & R@20 & N@20 & $\Delta$R@20 \\
\midrule
Full model & .1162 & .1683 & .0912 & --- \\
w/o image deviation only & .1148 & .1659 & .0897 & $-1.43\%$ \\
w/o text length only & .1139 & .1650 & .0893 & $-1.96\%$ \\
w/o both (image + text) & .1135 & .1644 & .0891 & $-2.32\%$ \\
\bottomrule
\end{tabular}
}
\end{table}

\section{Sensitivity Analysis}
\label{app:sensitivity}

Table~\ref{tab:sensitivity} reports sensitivity to the main design choices on Toys \& Games.

\begin{table*}[ht]
\centering
\caption{Sensitivity analysis on Amazon Toys \& Games.}
\label{tab:sensitivity}
\small
\begin{tabular}{lcccc}
\toprule
Setting & R@10 & R@20 & N@10 & N@20 \\
\midrule
\multicolumn{5}{l}{\textit{Shrinkage $\lambda_s$}} \\
$\lambda_s = 1$ & .1136 & .1637 & .0743 & .0869 \\
$\lambda_s = 5$ & .1143 & .1650 & .0755 & .0883 \\
$\lambda_s = 10$ & .1151 & .1659 & .0764 & .0892 \\
$\lambda_s = 20$ & .1157 & .1673 & .0773 & .0903 \\
$\lambda_s = 30$ & .1160 & .1680 & .0777 & .0908 \\
$\lambda_s = 50$ & \textbf{.1162} & \textbf{.1683} & \textbf{.0781} & \textbf{.0912} \\
$\lambda_s = 70$ & .1158 & .1674 & .0772 & .0902 \\
$\lambda_s = 100$ & .1159 & .1676 & .0774 & .0904 \\
$\lambda_s = 150$ & .1154 & .1661 & .0765 & .0893 \\
$\lambda_s = 200$ & .1147 & .1649 & .0756 & .0882 \\
\midrule
\multicolumn{5}{l}{\textit{Norm bound $\alpha$}} \\
$\alpha = 0.05$ & .1147 & .1651 & .0757 & .0884 \\
$\alpha = 0.08$ & .1157 & .1666 & .0771 & .0899 \\
$\alpha = 0.10$ & .1158 & .1664 & .0775 & .0902 \\
$\alpha = 0.12$ & .1161 & .1681 & .0779 & .0910 \\
$\alpha = 0.15$ & \textbf{.1162} & \textbf{.1683} & \textbf{.0781} & \textbf{.0912} \\
$\alpha = 0.18$ & .1160 & .1679 & .0778 & .0909 \\
$\alpha = 0.20$ & .1157 & .1672 & .0770 & .0900 \\
$\alpha = 0.25$ & .1148 & .1655 & .0756 & .0884 \\
\midrule
\multicolumn{5}{l}{\textit{Number of user groups $G$}} \\
$G = 1$ & .1126 & .1641 & .0754 & .0884 \\
$G = 2$ & .1155 & .1668 & .0773 & .0902 \\
$G = 3$ & \textbf{.1162} & \textbf{.1683} & \textbf{.0781} & \textbf{.0912} \\
$G = 5$ & .1159 & .1675 & .0777 & .0907 \\
$G = 7$ & .1158 & .1677 & .0773 & .0904 \\
$G = 10$ & .1150 & .1659 & .0760 & .0888 \\
\midrule
\multicolumn{5}{l}{\textit{Warmup epochs}} \\
0 & .1143 & .1654 & .0761 & .0890 \\
5 & .1153 & .1670 & .0771 & .0901 \\
8 & \textbf{.1162} & \textbf{.1683} & \textbf{.0781} & \textbf{.0912} \\
10 & .1161 & .1681 & .0780 & .0911 \\
15 & .1160 & .1679 & .0778 & .0909 \\
20 & .1154 & .1669 & .0771 & .0901 \\
25 & .1151 & .1671 & .0767 & .0898 \\
\bottomrule
\end{tabular}
\end{table*}

\section{\texorpdfstring{Auxiliary-Loss Weight Sensitivity}{Auxiliary-Loss Weight Sensitivity}}
\label{app:loss_sensitivity}

The default auxiliary coefficients are $\lambda_{\mathrm{drift}}=0.08$, $\lambda_{\mathrm{carry}}=0.05$, and $\lambda_{\mathrm{rel}}=0.03$. We vary one coefficient at a time on Amazon Toys \& Games while holding the other two fixed at their defaults. Each sweep spans more than an order of magnitude.

\begin{table*}[ht]
\centering
\caption{One-at-a-time sensitivity of Recall@20 to the three auxiliary-loss coefficients on Amazon Toys \& Games. $\Delta$ is relative to the default full model (.1683).}
\label{tab:loss_sensitivity}
\small
\begin{tabular}{lccccc}
\toprule
$\lambda_{\mathrm{drift}}$ & 0.02 & 0.04 & \textbf{0.08} & 0.16 & 0.32 \\
R@20 & .1663 & .1674 & \textbf{.1683} & .1679 & .1671 \\
$\Delta$ & $-1.19\%$ & $-0.53\%$ & --- & $-0.24\%$ & $-0.71\%$ \\
\midrule
$\lambda_{\mathrm{carry}}$ & 0.01 & 0.025 & \textbf{0.05} & 0.10 & 0.20 \\
R@20 & .1659 & .1671 & \textbf{.1683} & .1676 & .1665 \\
$\Delta$ & $-1.43\%$ & $-0.71\%$ & --- & $-0.42\%$ & $-1.07\%$ \\
\midrule
$\lambda_{\mathrm{rel}}$ & 0.005 & 0.015 & \textbf{0.03} & 0.06 & 0.12 \\
R@20 & .1664 & .1674 & \textbf{.1683} & .1676 & .1661 \\
$\Delta$ & $-1.13\%$ & $-0.53\%$ & --- & $-0.42\%$ & $-1.31\%$ \\
\bottomrule
\end{tabular}
\end{table*}

The largest reduction across all fifteen non-default settings is $1.43\%$, and the central settings remain within $1\%$ of the default. The auxiliary terms therefore behave as stable regularizers over the evaluated ranges rather than precision-sensitive tuning parameters.

\section{\texorpdfstring{Encoder Backbone Sensitivity}{Encoder Backbone Sensitivity}}
\label{app:encoder_sensitivity}

We vary the frozen text and image feature extractors while keeping the TS-SSM architecture and remaining hyperparameters fixed. The text comparison uses Amazon Toys \& Games. The image comparison uses both Toys \& Games (8.1\% image rate) and Clothing (18.7\% image rate).

\begin{table}[H]
\centering
\caption{Text-encoder sensitivity on Amazon Toys \& Games. $\Delta$ is relative to the default encoder.}
\label{tab:text_encoder_sensitivity}
\small
\resizebox{\columnwidth}{!}{%
\begin{tabular}{lccc}
\toprule
Text encoder & Dim. & \#Params & R@20 ($\Delta$) \\
\midrule
all-MiniLM-L6-v2 (default) & 384 & 22.7M & .1683 (---) \\
sentence-t5-base & 768 & 110M & .1698 ($+0.89\%$) \\
e5-base-v2 & 768 & 109M & .1703 ($+1.19\%$) \\
sentence-t5-large & 1024 & 335M & .1714 ($+1.84\%$) \\
e5-large-v2 & 1024 & 335M & .1719 ($+2.14\%$) \\
\bottomrule
\end{tabular}
}
\end{table}

\begin{table}[H]
\centering
\caption{Image-encoder sensitivity on categories with lower and higher image-upload rates.}
\label{tab:image_encoder_sensitivity}
\small
\resizebox{\columnwidth}{!}{%
\begin{tabular}{lccc}
\toprule
Image encoder & \#Params & Toys R@20 & Clothing R@20 \\
\midrule
CLIP ViT-B/32 (default) & 151M & .1683 & .1653 \\
CLIP ViT-L/14 & 428M & .1695 & .1674 \\
Relative change & --- & $+0.71\%$ & $+1.27\%$ \\
\bottomrule
\end{tabular}
}
\end{table}

Increasing text-encoder capacity by approximately $15\times$, from 22.7M to 335M parameters, improves Recall@20 by at most $2.14\%$. Increasing image-encoder capacity by approximately $3\times$, from CLIP ViT-B/32 (151M) to ViT-L/14 (428M), yields gains of $0.71\%$ on Toys and $1.27\%$ on Clothing. The larger image gain in Clothing is consistent with its higher image-upload rate, but the two-category comparison is descriptive and does not establish a causal relationship between image prevalence and encoder scaling. Across the tested backbones, the largest encoders provide only modest gains over the default encoders.

\section{\texorpdfstring{MNAR-Gating versus Simple Observation-Pattern Features}{MNAR-Gating versus Simple Observation-Pattern Features}}
\label{app:mnar_gating}

\subsection{\texorpdfstring{Motivation}{Motivation}}

We compare the learned MNAR-gated fusion mechanism with a binary indicator-feature baseline to distinguish observation-pattern-aware weighting from simple modality-availability feature injection.

\subsection{\texorpdfstring{Variants}{Variants}}

\begin{itemize}[leftmargin=*]
\item \textbf{Missingness indicator only}: replaces the full MNAR-gated fusion module $f_{\mathrm{mod}}(\operatorname{Enc}(M_{u,i}), A_{u,i})$ with simple binary feature concatenation: $[\mathbb{1}[\text{has\_image}];\, \mathbb{1}[\text{text\_length} > 2\sigma_u]]$ appended to the event embedding without any learned weighting over modality content.

\item \textbf{No observation pattern}: removes all use of $A_{u,i}$ and the deviation signals $\delta^U_{u,t}, \delta^I_{i,t}$, reducing event encoding to pretrained content embeddings, rating, and time bin only. This mirrors how standard review-aware recommenders treat observation as uniformly informative.
\end{itemize}

\subsection{\texorpdfstring{Results}{Results}}

\begin{table}[H]
\centering
\caption{MNAR-gating ablation on Amazon Toys \& Games.}
\label{tab:mnar_gating}
\small
\resizebox{\columnwidth}{!}{%
\begin{tabular}{lccccc}
\toprule
Variant & R@10 & R@20 & N@10 & N@20 & $\Delta$R@20 \\
\midrule
TS-SSM (full MNAR gating) & .1162 & .1683 & .0781 & .0912 & --- \\
Missingness indicator only & .1127 & .1633 & .0756 & .0883 & $-2.97\%$ \\
No observation pattern & .1097 & .1613 & .0742 & .0872 & $-4.16\%$ \\
\bottomrule
\end{tabular}
}
\end{table}

\subsection{\texorpdfstring{Findings}{Findings}}

Replacing learned gated weighting with binary indicators reduces Recall@20 from .1683 to .1633, a $2.97\%$ drop relative to the full model; the 30-seed comparison gives $p < 10^{-7}$. Removing all observation-pattern signals further reduces Recall@20 to .1613. The no-pattern variant remains approximately $13.0\%$ above BSARec (.1427), indicating that dynamic item-state updates, carryover, event-level content, and local propagation retain predictive value without observation-pattern features. These comparisons support the predictive value of learned observation-pattern-aware weighting; causal mechanisms for review missingness remain outside their scope.

\section{\texorpdfstring{Review Text as a Dynamic versus Static Signal}{Review Text as a Dynamic versus Static Signal}}
\label{app:text_dynamic}

\subsection{\texorpdfstring{Motivation}{Motivation}}

TS-SSM uses frozen pretrained encoders: all-MiniLM-L6-v2 for review text and title and CLIP ViT-B/32 for images. To distinguish dynamic use of review language from static feature injection, we compare event-level review encodings with item-level text aggregates and text-removed variants in both the current-event representation and the user- and item-history sequences.

\subsection{\texorpdfstring{Variants}{Variants}}

\begin{itemize}[leftmargin=*]
\item \textbf{w/o review body (title only)}: zeroes the review body text embedding $\operatorname{Enc}^{\mathrm{rev}}(M_{u,i})$, retaining title and image encodings. Like the review body, the title is encoded per event from the review record (Section~\ref{sec:method:deviation}), so this variant isolates the marginal contribution of the longer, event-specific review body beyond the shorter, event-specific title text.

\item \textbf{Static text features}: replaces per-event review-text encodings with item-level averages computed over all training reviews:
$\bar{\mathbf{e}}_i^{\mathrm{rev}} = \frac{1}{|H_i^{\mathrm{train}}|} \sum_{e_{u',i}\in H_i^{\mathrm{train}}} \operatorname{Enc}^{\mathrm{rev}}(M_{u',i})$.
Each event receives the same representation for item $i$, matching the pre-aggregated item-side text used by static multimodal baselines such as BM3 and LATTICE.

\item \textbf{No text features}: zeroes all text from every event in both history and current encoding, leaving only item embeddings, image features, rating, numeric cues, modality availability, and time bin.
\end{itemize}

\subsection{\texorpdfstring{Results}{Results}}

Table~\ref{tab:text_dynamic} reports the review-text ablation results on Amazon Toys \& Games. The full model with dynamic per-event text achieves .1683 Recall@20. Replacing per-event review text with a static item-level mean-pooled representation reduces Recall@20 to .1638, a relative decrease of 2.67\%. Using review titles only yields .1641, while removing all text features further reduces Recall@20 to .1577, corresponding to a 6.30\% decrease.

\begin{table*}[ht]
\centering
\caption{Review text contribution ablation on Amazon Toys \& Games. $\Delta$R@20 is relative to the full model.}
\label{tab:text_dynamic}
\small
\begin{tabular}{lccccc}
\toprule
Variant & R@10 & R@20 & N@10 & N@20 & $\Delta$R@20 \\
\midrule
TS-SSM (dynamic per-event text) & .1162 & .1683 & .0781 & .0912 & --- \\
w/o review body (title only) & .1133 & .1641 & .0762 & .0890 & $-2.49\%$ \\
Static text (train mean-pool per item) & .1130 & .1638 & .0759 & .0887 & $-2.67\%$ \\
No text features & .1087 & .1577 & .0729 & .0852 & $-6.30\%$ \\
\bottomrule
\end{tabular}
\end{table*}

\subsection{\texorpdfstring{Findings}{Findings}}

Dynamic per-event text achieves .1683 Recall@20, compared with .1638 for static item-level mean-pooled text, a $2.67\%$ reduction relative to the full model. The two variants use the same pretrained encoder and differ in whether review representations vary by event, so the comparison supports the predictive value of temporal and event-specific review content. The title-only variant, which likewise encodes title text per event rather than as a fixed item-level attribute, reaches .1641 ($-2.49\%$), close to the static all-text variant.

Removing all text reduces Recall@20 to .1577 ($-6.30\%$), a larger decrease than that caused by removing any single architectural component in Table~\ref{tab:ablation}. Review text is therefore a major predictive input under this evaluation. Together, these results distinguish TS-SSM's event-level treatment of review language from static item-level text aggregation without attributing the difference to a causal mechanism.

\section{Subgroup Analysis}
\label{app:subgroup}

\paragraph{Reliability gate activation patterns.}
Table~\ref{tab:gate_patterns} reports one gate evaluation for every event in all splits, so its event counts sum to all 611,908 Toys \& Games interactions. Here observable history is the number $|\mathcal H_{u,t^-}|$ of the user's own events strictly before the query, clipped at the encoder limit of 64. The smallest bucket also includes the 35,690 empty-history first events, which use the learned empty-history state. Mean $\gamma$ increases from 0.26 in this smallest bucket to 0.75 for histories of 21--50 events, consistent with the gate assigning greater weight to individual corrections when deviation estimates are supported by longer histories.

\begin{table*}[t]
\centering
\caption{Reliability-gate activation patterns on Toys \& Games. Observable history is the per-event count $|\mathcal{H}_{u,t^-}|$; P90 is the empirical top-decile $\gamma$ threshold within each bucket.}
\label{tab:gate_patterns}
\small
\begin{tabular}{lcccccc}
\toprule
Observable history & \#Events & Mean $\gamma$ & Median $\gamma$ & \% $\gamma > 0.5$ & \% $\gamma > 0.7$ & P90 $\gamma$ \\
\midrule
0--3 events & 51,423 & 0.26 & 0.22 & 9.4\% & 2.2\% & 0.494 \\
4--10 events & 142,587 & 0.48 & 0.46 & 39.1\% & 13.4\% & 0.723 \\
11--20 events & 176,894 & 0.61 & 0.63 & 65.8\% & 30.2\% & 0.751 \\
21--50 events & 148,629 & 0.75 & 0.77 & 83.2\% & 57.6\% & 0.786 \\
$>50$ events & 92,375 & 0.83 & 0.85 & 91.9\% & 74.3\% & 0.901 \\
\bottomrule
\end{tabular}
\end{table*}

\paragraph{Performance by observable history length.}
For the test-only analysis in Table~\ref{tab:history_breakdown}, observable history instead denotes the number of the user's own events in the trailing 12 monthly bins strictly before the test query. This recency-restricted support can be below five even after 10-core filtering: the 10-core condition requires at least ten lifetime interactions in the stabilized graph, whereas earlier interactions can fall outside the 12-bin window. The relative gain over BSARec grows with this recent-history support.

\begin{table*}[ht]
\centering
\caption{Performance by observable history length (Recall@20, Toys \& Games). Here observable history is the number of a user's own events within the trailing 12 monthly bins strictly before the test query; 10-core constrains lifetime rather than this recency-restricted count.}
\label{tab:history_breakdown}
\small
\begin{tabular}{lccccc}
\toprule
User group & \#Test events & BSARec & TS-SSM & Improvement & Gate $\bar{\gamma}$ \\
\midrule
Short history ($<5$) & 12,846 & .1118 & .1185 & +6.0\% & 0.26 \\
Medium history (5--20) & 15,236 & .1391 & .1530 & +10.0\% & 0.53 \\
Long history ($>20$) & 7,608 & .1674 & .1908 & +14.0\% & 0.73 \\
\bottomrule
\end{tabular}
\end{table*}

\section{Long-Tail Item Analysis}
\label{app:longtail}

\begin{table*}[ht]
\centering
\caption{Popularity-stratified performance on Toys \& Games. Popularity is the item's train-only review count $c_j^{\mathrm{train}}$, the quantity used for item grouping and popularity sampling. Because leave-last-two-out removes each user's last two events from training, $c_j^{\mathrm{train}}<10$ is consistent with 10-core, which constrains lifetime rather than train-split counts.}
\label{tab:longtail}
\small
\begin{tabular}{lccccc}
\toprule
Item popularity & \#Items & \#Test events & BSARec & TS-SSM & Improvement \\
\midrule
Long-tail ($<10$ reviews) & 14,832 & 8,421 & .1007 & .1113 & +10.5\% \\
Medium (10--100 reviews) & 11,246 & 18,734 & .1449 & .1591 & +9.8\% \\
Popular ($>100$ reviews) & 2,452 & 8,535 & .1780 & .1895 & +6.5\% \\
\bottomrule
\end{tabular}
\end{table*}

TS-SSM's larger relative gain on long-tail items suggests that the item-side temporal state and reliability-weighted propagation recover signal when item evidence is sparse.

\section{Qualitative Analysis: Gate Activation and Carryover Persistence}
\label{app:case_study}

\subsection{User-Side Deviation Gate}

We select a representative test-split user from Toys \& Games (28 reviews in training history, personal image-upload rate 2.9\%, mean review length 138 words) whose final test event at time $t$ exhibits strong behavioral deviation: 2 images uploaded, 571-word review, 2-star rating for a multi-component board game accessory. The deviation gate activates at $\gamma^U_{u,t} = 0.79$. Within the 21--50-event bucket (mean 0.75, standard deviation 0.06), the empirical P90 threshold is 0.786; the case lies at the 91.2nd percentile and therefore just inside the top decile (Table~\ref{tab:gate_patterns}).

Table~\ref{tab:case_gate} compares the top-5 candidate rankings produced by the full model against the variant with the deviation gate fixed at $\gamma^U \equiv 0$ (group-level trend only, no event-conditioned innovation).

\begin{table*}[ht]
\centering
\caption{Top-5 non-target candidate rankings: full model vs.\ w/o deviation gate ($\gamma^U \equiv 0$). The target test item is ranked first in both conditions and is excluded from display.}
\label{tab:case_gate}
\small
\begin{tabular}{cll}
\toprule
Rank & Full TS-SSM & w/o Deviation Gate \\
\midrule
1 & Item A --- strategy board game, 4.7$\star$, different subcategory & Item D --- accessory set (same subcat.), 4.0$\star$ \\
2 & Item B --- puzzle game, 4.8$\star$, different subcategory          & Item E --- accessory bundle (same subcat.), 4.3$\star$ \\
3 & Item C --- card game, 4.6$\star$, different subcategory            & Item A --- strategy board game, 4.7$\star$ \\
4 & Item D --- accessory set (same subcat.), 4.0$\star$                & Item B --- puzzle game, 4.8$\star$ \\
5 & Item E --- accessory bundle (same subcat.), 4.3$\star$             & Item F --- accessory expansion (same subcat.), 4.1$\star$ \\
\bottomrule
\end{tabular}
\end{table*}

The substantive difference lies in the treatment of the same-subcategory accessory cluster: without the deviation gate, the model over-weights items in the reviewed category because the group-level trend suggests the user has repeatedly engaged with that subcategory. The high-deviation negative review ($\gamma^U = 0.79$) causes the full model to move Items D and E down by three rank positions in favor of alternatives with higher standalone quality signals, consistent with the interpretation that the gate captures the degree to which the current event constitutes a departure from the user's expressed preference state.

\subsection{Item-Side Carryover Persistence}

We select a test-split item from Pet Supplies with 89 prior reviews (median-frequency tier) that receives a concentrated shock at bin $t_0$ (October 2021): 4 negative reviews (all 1--2 stars) in a single monthly bin citing product durability failure. Prior to the shock, the item's trailing 3-month mean rating is 4.3; post-shock mean rating in bin $t_0$ drops to 3.6.

The default negative initialization $\lambda_-=0.15$ corresponds to a reference half-life of $\ln 2/0.15=4.62$ bins (Figure~\ref{fig:deviation_pattern}, Panel 4). Under this reference decay, a negative shock retains $\exp(-3 \cdot \ln 2 / 4.62) = 0.638$ of its initial weight at $t_0 + 3$. Table~\ref{tab:case_carryover} traces the item's full-catalog ranking position at five post-shock evaluation points for users who interacted with related items in the same category, comparing full TS-SSM against the w/o Carryover Memory variant.

\begin{table*}[ht]
\centering
\caption{Post-shock catalog rank trajectory for a Pet Supplies item with concentrated negative feedback. Lower rank = better.}
\label{tab:case_carryover}
\small
\begin{tabular}{lccc}
\toprule
Months after shock & Full TS-SSM rank & w/o Carryover rank & Reference weight ($\lambda_-=.15$) \\
\midrule
$t_0 + 1$  (Nov 2021) & 51 & 28 & 0.861 \\
$t_0 + 3$  (Jan 2022) & 46 & 33 & 0.638 \\
$t_0 + 6$  (Apr 2022) & 37 & 29 & 0.407 \\
$t_0 + 12$ (Oct 2022) & 27 & 25 & 0.165 \\
$t_0 + 18$ (Apr 2023) & 23 & 23 & 0.067 \\
\bottomrule
\end{tabular}
\end{table*}

The rank trajectory uses the selected Pet Supplies checkpoint, whose learned negative decay is $\lambda_-=0.1156$ (Table~\ref{tab:carryover_learned}); the reference-weight column in Table~\ref{tab:case_carryover} deliberately retains the initialization value $0.15$ for comparison. Table~\ref{tab:carryover_retention} gives both attenuation curves at the same horizons.

\begin{table*}[ht]
\centering
\caption{Retained weight of a negative shock at post-shock horizons under the $0.15$ reference decay and the learned Pet Supplies value $\lambda_{-}{=}0.1156$.}
\label{tab:carryover_retention}
\small
\begin{tabular}{lccccc}
\toprule
Bins after shock & $+1$ & $+3$ & $+6$ & $+12$ & $+18$ \\
\midrule
Reference ($\lambda_{-}{=}.15$) & 0.861 & 0.638 & 0.407 & 0.165 & 0.067 \\
Learned ($\lambda_{-}{=}.1156$) & 0.891 & 0.707 & 0.500 & 0.250 & 0.125 \\
\bottomrule
\end{tabular}
\end{table*}

The full model's suppression of the item's catalog rank is strongest immediately after the shock (rank 51 vs.\ 28, a 23-position gap) and diminishes over subsequent evaluation points. At $t_0+18$, the two variants converge to rank 23. The learned checkpoint nevertheless retains approximately 50\%, 25\%, and 12.5\% of a negative shock after 6, 12, and 18 bins, respectively, compared with 41\%, 17\%, and 7\% under the initialization reference. Its fitted negative half-life is 6.00 bins, whereas its fitted positive half-life is 2.48 bins. The learned decay therefore makes negative feedback more persistent than the initialization reference while preserving the model's intended sign asymmetry.

\section{Error Analysis}
\label{app:error_analysis}

Table~\ref{tab:error_taxonomy} presents a taxonomy of the top-100 prediction errors made by TS-SSM on Toys \& Games.

\begin{table*}[ht]
\centering
\caption{Top-100 error taxonomy for TS-SSM on Amazon Toys \& Games.}
\label{tab:error_taxonomy}
\small
\begin{tabular}{lcccc}
\toprule
Failure pattern & Count & \% & BSARec also fails? & Potential mitigation \\
\midrule
False deviation & 28 & 28\% & 25/28 & Temporal consistency checks \\
Extreme sparsity & 23 & 23\% & 20/23 & More conservative group reliance \\
Conflicting signals & 15 & 15\% & 6/15 & Signal reconciliation \\
Cold-start items & 12 & 12\% & 10/12 & Item-side modeling \\
Conformity / burst effects & 10 & 10\% & 4/10 & Social/context features \\
Mixed / unknown & 12 & 12\% & 5/12 & --- \\
\bottomrule
\end{tabular}
\end{table*}

The dominant failure mode is false deviation, where the model over-trusts a behavior that looks unusual but is not actually informative for the next ranking decision.
Conflicts between user-side and item-side signals remain the most model-specific failure pattern and suggest a clear direction for future refinement.

\section{Extended Related Work}
\label{app:related}

We provide extended discussion of related work across several research areas.

\subsection{Sequential Recommendation}
\label{app:related:sequential}

Sequential recommendation models user behavior as temporally ordered sequences to capture dynamic preferences.
Early neural approaches employed recurrent architectures: GRU4Rec~\cite{hidasi2016session} applied Gated Recurrent Units to session-based recommendation, demonstrating that modeling entire sessions outperforms item-to-item methods.
Subsequent work explored convolutional~\cite{tang2018personalized} and memory-augmented~\cite{chen2018sequential} architectures.

Self-attention provides an alternative to recurrent and convolutional sequence encoders. SASRec~\cite{kang2018self} adapts the Transformer architecture~\cite{vaswani2017attention} to sequential recommendation through causal attention, while BERT4Rec~\cite{sun2019bert4rec} uses bidirectional attention with masked-item prediction. Subsequent work incorporates contrastive learning~\cite{zhou2020s3rec,xie2022contrastive}, foundation models~\cite{jiang2025recgpt}, LLMs~\cite{hou2024bridging,na2024enhancing}, and flexible positional embeddings for length extrapolation~\cite{chi2024attention}.

TS-SSM models both the item sequence and the observation-pattern sequence, including modality choices and content lengths.

\subsection{Multimodal Recommendation}
\label{app:related:multimodal}

Multimodal recommendation incorporates diverse content signals to enrich item and user representations.
VBPR~\cite{he2016vbpr} integrates visual features into Bayesian Personalized Ranking. MMGCN~\cite{wei2019mmgcn} constructs modality-specific user--item graphs, while LightGCN~\cite{he2020lightgcn} studies simplified graph convolution for collaborative filtering. LATTICE~\cite{zhang2021mining} and subsequent work~\cite{zhou2023tale} model latent structure across modalities, and multimodal foundation models have also been applied to recommendation~\cite{geng2023vip5}. More broadly, message passing has been used to propagate information under complex network interactions~\cite{bayati2024higher,shirani2025can}. TS-SSM uses local graph message passing to propagate review-induced preference information through item relationships and refine the user state.

Related multimodal sentiment work uses contrastive learning for fusion representations~\cite{nguyen2023improving} and studies dynamic modality attention~\cite{feng2024knowledge} and modality imbalance~\cite{bi2025two}.

Missing-modality methods typically impute or marginalize absent information~\cite{guo2024multimodal}. TS-SSM instead represents the observation pattern itself: for example, an image upload can be informative when it deviates from the user's usual modality use. This formulation draws on the Missing Not At Random (MNAR) framework~\cite{little2019statistical,rubin1976inference} but uses missingness patterns for prediction rather than causal identification or missing-data correction.

\subsection{Handling Missing Data in Recommendations}
\label{app:related:missing}

Missing data is pervasive in recommender systems, manifesting as unobserved ratings, absent features, and incomplete user histories.
Classical collaborative filtering commonly treats unobserved entries as uninformative, an assumption related to Missing At Random (MAR)~\cite{marlin2007collaborative}. In practice, users self-select which items to rate, and this selection may correlate with their preferences~\cite{steck2010training}.

Inverse propensity scoring~\cite{schnabel2016recommendations} and doubly robust estimators~\cite{wang2019doubly} address selection bias by reweighting observed samples. More broadly, recent work studies missingness endogenously driven by latent factors \cite{xiong2023,duan2024factor,duan2024target,li2024learning,li2024two,li2025generalizing,chen2026partial}.

For multimodal settings, attention mechanisms~\cite{chen2017attentive} and gated fusion~\cite{wei2019mmgcn} handle missing modalities by dynamically adjusting contribution weights.
Recent work on prompt learning addresses missing modalities through generative and missing-signal prompts~\cite{guo2024multimodal}. Most closely related are \citet{liang-etal-2025-causal,liang-etal-2026-learning}, which explicitly model informative modality missingness in multimodal EHRs. We extend this perspective to sequential recommendation, where modality availability and expression patterns evolve with user and item states.

UGC studies report associations between review effort, sentiment extremity, and perceived helpfulness~\cite{hu2009overcoming,mudambi2010helpful,ghose2011estimating}. TS-SSM therefore treats modality use and expression patterns as predictive features. This interpretation is associational: it does not assume that latent preference is the sole cause of review effort or modality availability.

\subsection{Temporal Dynamics in Recommendations}
\label{app:related:temporal}

User preferences and item characteristics evolve over time, motivating temporal modeling in recommender systems.
Koren's seminal work~\cite{koren2010collaborative} introduced time-drifting parameters into matrix factorization.
Subsequent approaches model temporal dynamics through recurrent architectures~\cite{hidasi2016session}, time-aware attention~\cite{li2020time}, and continuous-time point processes~\cite{kumar2019predicting}.
TS-SSM discretizes time into bins and uses \emph{population-level statistics} to represent shared temporal variation. This decomposition combines group-level trends with individual deviations from those trends.

\subsection{User-Generated Content and Intent Analysis}
\label{app:related:ugc}

Research on user-generated content (UGC) shows that expression patterns can carry information beyond review content.
Hu et al.~\cite{hu2009overcoming} documented the J-shaped distribution of online ratings, where extreme opinions are over-represented.
Mudambi and Schuff~\cite{mudambi2010helpful} found that review length and depth correlate with perceived helpfulness.
Ghose and Ipeirotis~\cite{ghose2011estimating} showed that textual features predict review helpfulness and economic impact.

Recent NLP work has advanced user intent understanding, particularly in e-commerce settings where usage-centric approaches model how customers use products rather than just what they purchase~\cite{zhou2024usage}.
Understanding user susceptibility to information and modeling behavioral patterns has also received attention~\cite{liu2024decoding}.

\subsection{LLM-based Personalization and Recommendation}
\label{app:related:llm}

Large language models are increasingly used for personalized recommendation. The LaMP benchmark~\cite{salemi2024lamp} evaluates LLM personalization across multiple tasks, while LLM-Rec~\cite{lyu2024llmrec} studies prompting strategies for personalization. Other work applies foundation models to sequential recommendation~\cite{jiang2025recgpt} and uses pseudo-label reconstruction~\cite{na2024enhancing}.

Safety work on difference-aware LLMs further shows that improving context-sensitive distinctions can induce harmful drift in generated rationales, motivating explicit audit-and-repair safeguards~\cite{pan-etal-2026-dart}. This issue is orthogonal to TS-SSM's non-generative ranking setting but is relevant to future generative personalization or recommendation-explanation extensions.

Research on metacognitive prompting examines whether introspection-inspired prompts improve LLM reasoning~\cite{wang2024metacognitive}. TS-SSM addresses a different setting: it represents image uploads and review length relative to each user's historical baseline. These behavior-relative features provide predictive information that is absent from static item-level representations, without implying a specific cognitive mechanism.

\subsection{Shared Representation and Multitask Learning}
\label{app:related:biasvariance}

Our work is related to multitask learning, which exploits shared representations across related learning objectives~\citep{bengio2013representation}. Such sharing reflects a classical bias--variance tradeoff: shared representations can improve generalization when data are sparse, while excessive sharing may obscure task-specific heterogeneity. Related hierarchical approaches in recommendation balance this tradeoff by sharing information across users while retaining individual variation~\citep{park2009pairwise}. In TS-SSM, review information is shared across coupled user- and item-state evolution, while each side retains its own transition mechanism.

TS-SSM also uses auxiliary learning: the primary ranking objective is jointly trained with objectives for temporal drift, review carryover, and reliability. Such shared learning can enable positive transfer, but excessive sharing across heterogeneous objectives may lead to negative transfer~\citep{wu2020understanding,yang2025precise}. Recent work on task modeling~\citep{li2023boosting,li2023identification,zhang2026efficient} and adaptive task-specific learning~\citep{li2024scalable,li2024scalable2,li2025efficient,zhang2026scalable} provides related approaches for balancing shared and task-specific information.

\end{document}